\documentclass[preprint2]{aastex6}
\usepackage{epsfig}
\usepackage{amsfonts,amssymb,amsmath}

\begin{document}

\title{Constraints on the Evolution of the Galaxy Stellar Mass Function I: Role of Star Formation, Mergers and Stellar Stripping}
\author{E. Contini$^{1}$, Xi Kang$^{1}$, A.D. Romeo$^{1}$, Q. Xia$^{1}$}
\affil{$^1$Purple Mountain Observatory, the Partner Group of MPI f\"ur Astronomie, 2 West Beijing Road, Nanjing 210008, China}

\email{contini@pmo.ac.cn, 
\\
\\
kangxi@pmo.ac.cn}

\begin{abstract} 
We study the connection between the observed star formation rate-stellar mass (SFR-$M_*$) relation and the evolution of 
the stellar mass function (SMF) by means of a Subhalo Abundance Matching technique coupled to merger trees extracted 
from a N-body simulation. Our approach, which considers both galaxy mergers and stellar stripping, is to 
force the model to match the observed SMF at redshift $z>2$, and let it evolve down to the present time according to 
the observed (SFR-$M_*$) relation. In this study, we use two different sets of SMFs and two SFR-$M_*$ relations: a simple 
power law and a relation with a mass-dependent slope.
Our analysis shows that the evolution of the SMF is more consistent with a SFR-$M_*$ relation with 
a mass-dependent slope, in agreement with predictions from other models of galaxy evolution and recent observations. In order 
to fully and realistically describe the evolution of the SMF, both mergers and stellar stripping must be considered, and we find that both have 
almost equal effects on the evolution of SMF at the massive end. Taking into account the systematic uncertainties in the 
observed data, the high-mass end of the  SMF obtained by considering 
stellar stripping results in good agreement with recent observational data from the Sloan Digital Sky Survey (SDSS). At 
$\log M_* < 11.2$, our prediction at z=0.1 is close to \citet{li-white09} data, but the high-mass end ($\log M_* > 11.2$) is 
in better agreement with \citet{dsouza15} data which account for more massive galaxies. 

\end{abstract}


\section[]{Introduction} 
\label{sec:intro}
 
An important goal of galaxy formation and evolution modelling is to understand the 
buildup of the stellar matter content in the Universe. This is achievable when observations 
at different redshifts can help models in constraining fundamental statitistics, such as the 
stellar mass function (SMF) and its evolution with time. On one hand, the SMF describes  
the galaxy stellar mass distribution that we can directly observe, and on the other hand, its
evolution with time provides a natural test for studying the physical processes responsible for 
the growth of galaxies. Therefore, the SMF can be used to constrain cosmological models and 
the physical processes taking place during galaxy formation. Recently, the redshift evolution 
of the SMF has been measured by several wide surveys (e.g., 
\citealt{perez-gonzalez08,drory09,marchesini09,santini12,moustakas13,muzzin13,ilbert13,tomczak14}; \citealt{tomczak16}).
The FourStar Galaxy Evolution Survey (ZFOURGE, \citealt{tomczak14}) is the latest example of a wide 
and deep near-infrared survey, with a mass completeness around $\log M_* \simeq 9.0$ at 
redshift $z\simeq 2$. Moreover, the SMF has been studied separately for the star-forming and quiescent 
populations (see, e.g., \citealt{drory09,tomczak14}). The change of the slope with time appears to be 
caused by the presence of diverse galaxy populations that lead to a mass function shape more complex than 
a single power law with an exponential cutoff (\citealt{drory09}). The need of a double-Schechter (\citealt{schechter76}) function
at $z<2$, rather than a single-Schechter, has been pointed out by other authors (e.g. \citealt{tomczak14}).

Another important statistical property is the relation between the star formation rate (SFR) of galaxies 
versus their stellar mass (SFR-$M_*$ relation, also called ``\emph{main sequence relationship}" (MS)), which 
describes the rate of star formation for galaxies with given stellar mass. The SFR-$M_*$ relation, generally 
assumed to be a linear relation in a log-plane, has been intensely examined during the recent past 
(\citealt{daddi07,elbaz07,rodighiero11,whitaker12,whitaker14,shivaei15,tasca15,tomczak16}). Overall these 
studies found an associated (due to observational uncertainties) and intrinsic scatter. While the scatter of 
this relation is found to remain tight at least up to $z\sim 1.5$ (\citealt{salmi12}), its slope is reported 
to vary between 0.6 and 1, depending mainly on sample definitions and star formation indicators. According to 
several works, a log-linear relation between SFR and $M_*$ gets bent towards of lower SFRs for higher stellar masses 
(e.g., \citealt{karim11,whitaker14,tomczak16}). Moreover, it has been established that the normalisation of the 
SFR-$M_*$ relation is redshift-dependent (\citealt{brinchmann04,elbaz07,daddi07,noeske07,renzini09,peng10,lee15}), 
while much more controversial appears the redshift evolution of its slope.

The SMF and SFR-$M_*$ relation are independent observables, and in principle it is 
possible to combine the observed SMF at high redshift with the SFR-$M_*$ 
relation to obtain subsequent stellar mass functions of galaxies. Such an approach has been 
followed by several authors (e.g., \citealt{conroy09,leja15,tomczak16}) in order to shed light on the 
connection between the two observed statistics. In fact, all studies so far have 
found that the two observations are inconsistent with the observed evolution 
of the SMF (Leja et al. 2015). How to self-consistently describe the relationship between these 
two statistics is currently a hot issue in the study of galaxy evolution. 

\cite{leja15} followed the aforementioned approach by connecting the observed star-forming 
sequence and the observed evolution of the SMF, between $0.2 < z < 2.5$. Interestingly, their
study suggests that a single slope of the star-forming sequence smaller than 0.9 at all 
masses and redshifts would result in a higher number density of intermediate stellar mass 
galaxies, close to the knee of the SMF. To alleviate the discrepancy, they suggest a broken power 
law with a shallower slope at high masses and find that such a SFR-$M_*$ relation better agrees with 
the SMF evolution, although the inferred SMF is still offset by 0.3 dex from the observed one. The 
result of their study clearly shows that a mass-dependent slope of the SFR-$M_*$ relation is then
needed to reconcile the observed evolution of the SMF with that inferred by connecting the SFR-$M_*$ 
relation with the observed SMF at high-redshift. According to their study, neither mergers nor hidden 
low-mass quiescent galaxies not detected are likely to be responsible for the mismatch. 

Similar conclusions have been reached in a very recent study by \cite{tomczak16}. These authors 
took advantage of the ZFOURGE survey to examine the slope of the SFR-$M_*$ relation. They find 
a redshift-dependent relation not consistent with a single power-law. The slope becomes shallower 
above a given turnover mass that ranges from $10^{9.5}-10^{10.8} M_{\odot}$. They use the
evolving SFR-$M_*$ relation with redshift to generate star-fomation histories of galaxies. By 
integrating the set of star-formation histories with time, they obtain mass-growth histories to 
compare to the mass growth from the evolution of the stellar mass function of \cite{tomczak14}. 
They find a reasonable agreement between the observed and inferred SMFs, but also room for a more detailed 
investigation. According to their conclusions, their study implies that either the star-formation 
rates measurements are overestimated, or the growth of the \cite{tomczak14} mass function is too 
slow, or both.

Here, and in a forthcoming paper (Contini et al. in prep.), we want to address this issue, following 
a similar method such as those described above. In this study, we adopt a relatively new approach. We use 
accurate merger trees constructed from N-body simulations of \cite{kang12}, and set up the stellar 
mass of galaxies by using an abundance matching technique in order to match the observed SMF at high
redshift. Thus, our initial SMF matches perfectly with the observed data. Then, we use two different 
flavours of the observed SFR-$M_*$ relation, a single power-law and a mass-dependent slope, to assign SFRs
to galaxies at any given lower redshift. We let galaxies grow according to their star-formation 
histories given by the SFR-$M_*$ relation as a function of redshift, and via mergers. This method is 
in spirit similar to \cite{conroy09}, but we proceed in the opposite direction, trying to reconcile the 
evolution of the SMF coupled to the observed SFR-$M_*$ relation, rather than guessing the latter 
from the former. Compared to previous studies, our novelty lies in two aspects: a) we take advantage 
of real merger trees from N-body simulations, which provide us the accretion histories of galaxies;
b) we include stellar stripping in the model.
Our main goals are:
\begin{itemize}
 \item [1)] understand the role of the SFR-$M_*$ relation (power-law or mass-dependent slope);
 \item [2)] understand the role of stellar stripping and mergers in the evolution of the SMF.
\end{itemize}

The paper is structured as follows: in Section \ref{sec:method} we describe in detail our approach 
to address the issue and the method followed. In Section \ref{sec:results} we show our results, which 
will be fully discussed in Section \ref{sec:discussion}. Finally, in Section \ref{sec:conclusions} we 
give our conclusions.

\section[]{Method}  
\label{sec:method}

The simulation used in this paper is based on \cite{kang12}. We refer the readers to that paper for more 
details, while here we introduce the main prescriptions. The simulation was performed using the Gadget-2 code 
(\citealt{springel05}) with cosmological parameters adopted from the WMAP7 data release (\citealt{komatsu}), 
namely: $\Omega_{\lambda}=0.73, \Omega_{m}=0.27, \Omega_{b}=0.044$, $h=0.7$ and $\sigma_{8}=0.81$. The simulation 
box is $200 \, Mpc/h$ on each side using $1024^{3}$ particles, each with mass $5.64\cdot 10^8 \, M_{\odot}h^{-1}$. 
The merger trees are constructed by following the subhaloes resolved in FOF haloes at each snapshot (e.g., 
\citealt{kang05}) making use of the algorithm SUBFIND (\citealt{springel01}). 

To populate dark matter haloes with galaxies we use the so-called Subhalo Abundance Matching (ShAM) technique. 
This method, originally proposed by \cite{vale04}, is now widely-used in numerical simulations to connect 
galaxies with dark matter structures
(\citealt{vale06,conroy06,behroozi10,moster10,guo10,hearin13,guo14,yamamoto15,chaves-montero16}). 
The fundamental assumption relies on a monotonic mapping between a given galaxy property,
typically luminosity or stellar mass, and a given property of subhaloes such as maximum mass or 
maximum circular velocity of the subhalo during its history. One advantage of this method is that it 
is relatively easy to use, because it requires very few assumptions and avoids the explicit modelling
of the physics of galaxy formation. Unfortunately, this entails a loss of a large amount of
information, that is, indeed, the main handicap of the modelling.

For the purpose of our study we need to connect galaxies and haloes at different times. We start by 
forcing the algorithm to match the observed SMF (see Section \ref{sec:smf} for more details concerning the 
SMFs chosen) at a given redshift. The algorithm makes use of the merger-trees described above and sorts galaxies 
and haloes in mass. Finally, it links them in a one-to-one relation. As time passes by, new haloes can form. 
We populate them with galaxies having stellar masses in accordance with the stellar mass-halo mass 
relation given by the abundance matching technique applied at that redshift. From the redshift of the 
match ($z_{match}$) we let galaxies evolve according to the their merger histories (given by the merger trees) 
and to their star formation histories. At each redshift we assign SFRs to galaxies by means of the SFR-$M_*$ 
relation observed at that redshift (see Section \ref{sec:smf}), down to $z=0$. In order to consider also tidal 
interactions between satellite galaxies and the potential well of their hosts, we implement in the model a 
simple prescription for stellar stripping (see Section \ref{sec:stripping} for details).

\subsection[]{Stellar Mass Functions and SFR-$M_*$ relations}  
\label{sec:smf}

We use two sets of observed stellar mass functions and two different SFR-$M_*$ relations, a single power-law 
and a mass-dependent slope relation. The first set of SMFs is taken from \cite{fontana06}, who use data 
from the GOODS-MUSIC catalog, from $0.4<z<4$. This catalog contains 2931 Ks-selected galaxies with multi-wavelength 
coverage extending from the U-band to the Spitzer 8 $\mu$m band, of which 27 per cent have spectroscopic 
redshifts and the remaining fraction have accurate photometric redshifts. For this sample they apply a standard 
fitting procedure to measure stellar masses, and compute the SMF up to $z\simeq 4$. These authors use a Salpeter 
(\citealt{salpeter55}) Initial Mass Function (IMF), while our algorithm makes use of a Chabrier IMF (\citealt{chabrier03}). 
Following \cite{longhetti09}, we have corrected their masses in order to be consistent with a Chabrier IMF by means of 
the following relation:
\begin{displaymath}
M^* _{Cha}(z)=0.55 \cdot M^* _{Sal} (z). 
\end{displaymath}

The second set of SMFs has been constructed by \cite{tomczak14}, using observations from the FourStar 
Galaxy Evolution Survey (ZFOURGE). These data represent the deepest measurements to date of the galaxy SMF 
in the redshift range $0.2 < z < 3$. ZFOURGE is composed of three $11' \times 11'$ pointings with coverage in the CDFS
(\citealt{giacconi02}), COSMOS (\citealt{capak07}) and UKIDSS (\citealt{lawrence07}). The ZFOURGE fields also 
take advantage of HST imaging taken as part of the CANDELS survey (\citealt{grogin11,koekemoer11}) and 
from the NEWFIRM Medium-Band Survey (NMBS; \citealt{whitaker11}).

Both sets of data have been fit with a single-Schechter function, defined as:
\begin{equation}
 \frac{\Phi(M)dM}{\ln(10)} = \phi^* \left[10^{(M-M^* )(1+\alpha^* )}\right]\exp \left( -10^{(M-M^* )}\right) dM
\end{equation}
where $M = \log(M/M_{\odot})$, $\alpha^*$ is the slope at the low-mass end, $\phi^*$ is the normalization and $M^*$ is 
the characteristic mass. For the set of SMFs by Fontana et al. the three parameters $\alpha^*$, $\phi^*$ and $M^*$
evolve with redshift according to the following set of parameterisations:
\begin{displaymath}
\phi^* (z) = \phi_0 ^* \cdot (1+z)^{\phi_1 ^*}
\end{displaymath}
\begin{equation}
\label{eq:parameterisation_font}
\alpha^* (z) = \alpha^* _0 + \alpha^*_1 \cdot z 
\end{equation}
\begin{displaymath}
M^* (z) = M^* _0 +M^* _1 \cdot z +M^* _2 \cdot z^2
\end{displaymath}
where $\phi^* _0$, $\phi^* _1$, $\alpha^* _0$, $\alpha^* _1$, $M^* _0$, $M^* _1$ and $M^* _2$ are free parameters which
values are reported in Table \ref{tab:fontana}. In Table \ref{tab:tomczak} we report the values of the three parameters 
of the fits at different redshifts for Tomczak et al.'s set. 

\begin{table}
\caption{Best fit of the seven free parameters in the set of equations \ref{eq:parameterisation_font}. Note that $M^*_0$
has been corrected for the different IMF. The original value in Fontana et al. is $M^*_0 = 11.16$.} 
\centering
\begin{tabular}{cc}
\hline 
$M^*_0$           & $10.90$       \\
$M^*_1$       & $+0.17\pm0.05$    \\ 
$M^*_2$            & $-0.07\pm0.01$ \\
$\alpha^*_0$   & $-1.18$   \\
$\alpha^*_1$  & $-0.082\pm0.033$  \\
$\phi^*_0$         & $0.0035$          \\
$\phi^*_1$      & $-2.20\pm0.18$     \\
\hline 
\end{tabular}
\label{tab:fontana}
\end{table}

\begin{table}
\caption{Redshift, logarithm of the normalisation, slope at the low-mass end and logarithm of the characteristic 
mass for the set of SMFs by Tomczak et al..}
\begin{center}
\begin{tabular}{llllll}
\hline
Redshift & $\log(\phi^* )$ & $\alpha^*$ & $\log(M^*)$ \\
\hline
 0.3 & $-2.96\pm0.10$  & $-1.35\pm0.04$ & $11.05\pm0.10$ \\
 0.6 & $-2.93\pm0.07$  & $-1.35\pm0.04$ & $11.00\pm0.06$ \\
 0.8 & $-3.17\pm0.11$  & $-1.38\pm0.04$ & $11.16\pm0.12$ \\
 1.1 & $-3.19\pm0.11$  & $-1.33\pm0.05$ & $11.09\pm0.10$ \\
 1.3 & $-3.11\pm0.08$  & $-1.29\pm0.05$ & $10.88\pm0.05$ \\
 1.8 & $-3.28\pm0.08$  & $-1.33\pm0.05$ & $11.03\pm0.05$ \\
 2.2 & $-3.59\pm0.14$  & $-1.43\pm0.08$ & $11.13\pm0.13$ \\
\hline
\end{tabular}
\end{center}
\label{tab:tomczak}
\end{table}

In order to account for star formation we use two SFR-$M_*$ relations: 1) a single power-law 
with normalisation redshift-dependent, and 2) a parameterised relation with a mass-dependent slope.
The single power-law relation reads as follow:
\begin{equation}\label{eq:single_slope}
 SFR \, [M_{\odot}/yr] = 2.78\cdot  M_{*,10}^{0.9} \cdot(1+z)^{1.8} \, ,
\end{equation}
where $M_{*,10}$ is the stellar mass in unit of $10^{10} \, M_{\odot}$. Equation \ref{eq:single_slope} 
is a good representation of the SFR-$M_*$ relations suggested by \cite{daddi07} at $z\sim 2$, and 
\cite{elbaz07} at $z\sim 1$. For the relation with a mass-dependent slope we have chosen the 
parameterisation suggested by \cite{tomczak16} (see also \citealt{lee15}):
\begin{equation}\label{eq:dep_slope}
  \log(SFR \, [M_{\odot}/yr]) = s_0 -\log \left[ 1+\left( \frac{M_*}{M_0}\right)^{\gamma} \right] \, ,
\end{equation}
where $s_0$ and $M_0$ are in units of $\log(M_{\odot}/yr)$ and $M_{\odot}$ respectively. Equation 
\ref{eq:dep_slope}, as explained by \cite{tomczak16}, behaves as a normal power-law with slope $\gamma$
at low masses, and asymptotically approaches a peak value $s_0$ above a characteristic stellar mass $M_0$.
Using data from the same survey used in \cite{tomczak14}, these authors find that such a parameterisation 
works well even if quiescent galaxies are considered. For this reason, we parameterise $s_0$ and $M_0$ with 
the same second-order polynomials valid for all galaxies (see Eq. 3 in \citealt{tomczak16}):
\begin{displaymath}
s_0 = 0.195  +  1.157 z  -  0.143 z^2
\end{displaymath}
\begin{equation}
\label{eq:parameterisation_all}
 \log(M_0 )= 9.244  +  0.753 z  -  0.090 z^2  
\end{equation}
\begin{displaymath}
 \gamma = -1.118
\end{displaymath}
Equation \ref{eq:dep_slope} and the set of equations \ref{eq:parameterisation_all} describe the evolution 
with time of the SFR-$M_*$ relation with a mass-dependent slope.

\subsection[]{Stellar Stripping}  
\label{sec:stripping}

Stellar stripping is an important process that takes place during galaxy formation and evolution.
It has been shown by several authors (e.g. \citealt{purcell07,myself2}) that most of the diffuse
light around galaxies forms through this channel. As highlighted in Section \ref{sec:intro}, no 
study that used a similar approach as ours considered the effect of stellar stripping on the evolution of the SMF, especially its role 
at redshift $z<1$, when the bulk of the diffuse light starts to form (\citealt{giuseppe,myself2}).
We model stellar stripping in a very simple fashion, assuming that the stellar mass lost due to 
disruption events and stripping can be parameterised by an exponential law such as:
\begin{equation}
 M_{lost}^{z=z_i} = M_{*(s1,s2)}^{z=z_i} \cdot \left( 1-\exp \left(\frac{-\eta (M_{halo})(t_{infall}-t_i)}{\tau_{s1,s2}}\right) \right) \, ,
\end{equation}
where $t_{infall}$ is the lookback time when the galaxy last entered a cluster (i.e., became a satellite),
$t_i$ the lookback  time at $z=z_i$, and $\tau_{s1,s2}$ two normalisations arbitrarily set to 30 Gyr and 15 Gyr, respectively,
for satellites associated with a distinct dark matter substructure ($s_1$), and orphan galaxies 
($s_2$). We set an higher normalisation for satellites $s_1$ in order to 
consider the effect of dark matter in shielding the galaxy from tidal forces \footnote{The evidence for dark matter shielding comes mainly from numerical
simulations. \cite{villalobos12} showed that, at odds to what usually assumed in semi-analytic models (which basically used to treat stellar stripping
with recipes involving orphan galaxies only), satellites associated with a subhalo can feel tidal forces and be subject to stellar stripping, even though in a less 
efficient way. \cite{myself2} considered these results and apply stripping to satellite with their own subhalo if and only if the half-mass radius of their parent 
subhalo is smaller than the half-mass radius of the galaxy's disk (equations 5 and 6 in \citealt{myself2}).}.

$\eta (M_{halo})$ is 
the stripping efficiency, chosen to be a function of the main halo mass ($M_{200}$ \footnote{$M_{200}$ is 
the mass within the virial radius $R_{200}$, defined as the radius that encloses a mean density of 200 times the
critical density of the Universe at the redshift of interest.}) in which satellites reside 
in order to consider the different strength of stellar stripping in haloes of different mass. We report 
in Table \ref{tab:eta_values} the values of $\eta$ for different halo mass ranges. These values have 
been chosen in order to reproduce as much as possible the observed abundance of satellites in groups and 
clusters (see Appendix \ref{sec:append}) by \citet{yang09}.

In addition to stellar stripping, we consider another channel for the formation of the diffuse light, 
i.e. the so called "merger channel" (see e.g., \citealt{giuseppe,myself2}), to consider also those 
stars that might end up in the diffuse light getting unbound through relaxation processes that happen 
during galaxy-galaxy merging. We simply assume that when two galaxies merge, 30 per cent of the satellite 
stellar mass gets unbound and goes to the diffuse component.

As stated above, the use of merger trees provided by our N-body simulation is one of the novelties of 
this work. Other authors, e.g. \cite{tomczak16}, take into account galaxy mergers and find that mergers  of small galaxies help to 
alleviate the discrepancy between the observed and predicted SMF, but it would require between $25-65$ per cent
of small galaxies to merge with a more massive galaxy per Gyr, a merger rate which highly exceeds current estimates of
galaxy merger rates. We compared such a high merger rate with that predicted by our model and found similar rates 
as those found by previous works (e.g. \citealt{lotz11,williams11,leja15}).

\begin{table}
\caption{Stripping efficiency $\eta$ (second line) as a function of the halo mass $M_{200}$ in unit of 
$M_{\odot}/h$ (first line).}
\begin{center}
\begin{tabular}{llllll}
\hline
 $<10^{13}$ & [1-5]$\cdot 10^{13}$ & [5-10]$\cdot 10^{13}$ & [1-5]$\cdot 10^{14}$ & [5-10]$\cdot 10^{14}$ & $>10^{15}$ \\
\hline
1 & 0.8 & 0.5 & 0.3 & 0.2 & 0.1\\
\hline
\end{tabular}
\end{center}
\label{tab:eta_values}
\end{table}

\section{Results}
\label{sec:results}

In this section we show our model predictions compared with observed data, the two sets of SMFs presented in Section \ref{sec:smf}. 
For the sake of simplicity, we define:
\begin{itemize}
 \item Model A: Fontana+06 SMFs coupled to the SFR-$M_*$ relation given by Equation \ref{eq:single_slope};
 \item Model B: Tomczak+14 SMFs coupled to the SFR-$M_*$ relation given by Equation \ref{eq:single_slope};
 \item Model C: Fontana+06 SMFs coupled to the SFR-$M_*$ relation given by Equation \ref{eq:dep_slope} and parameterisations given by 
  the set of equations \ref{eq:parameterisation_all};
 \item Model D: Tomczak+14 SMFs coupled to the SFR-$M_*$ relation given by Equation \ref{eq:dep_slope} and parameterisations given by 
  the set of equations \ref{eq:parameterisation_all}.
\end{itemize}
Unless otherwise specified, all models will take into account galaxies mergers and stellar stripping (see Section \ref{sec:stripping}).

\begin{figure*} 
\begin{center}
\includegraphics[scale=.6]{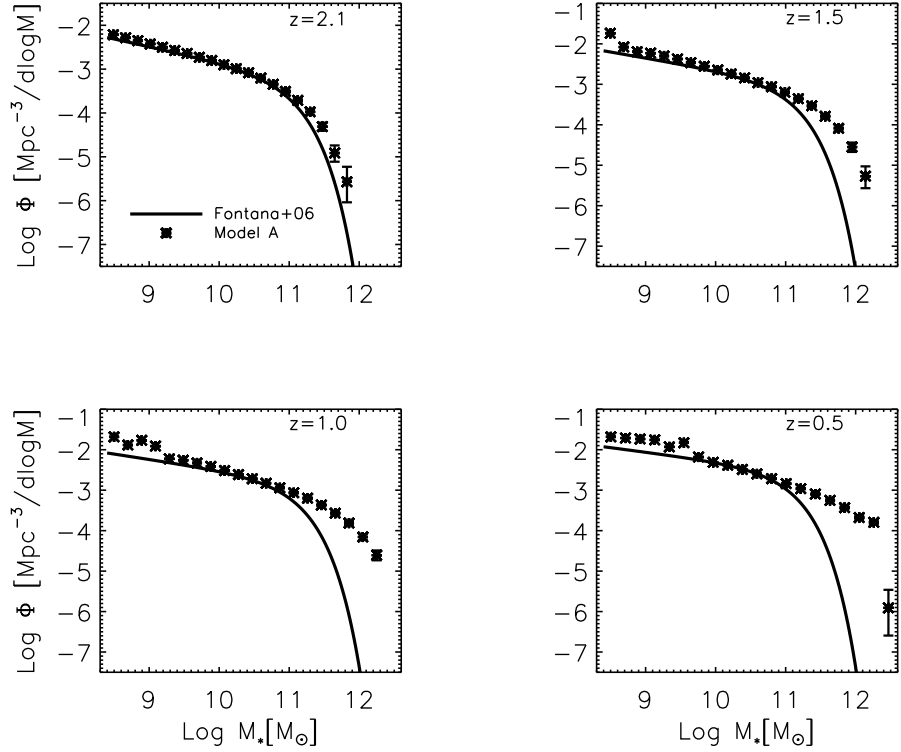} 
\caption{Evolution of the SMF from $z=2.1$ down to $z=0.5$ predicted by our Model A. Here we assume an initial SMF by Fontana+06 
at $z=2.6$ and coupled to a SFR-$M_*$ relation with a single slope. Stars represent model data while the solid lines represent 
Fontana+06 fit.}
\label{fig:fontana_ss}
\end{center}
\end{figure*}

\begin{figure*} 
\begin{center}
\includegraphics[scale=.6]{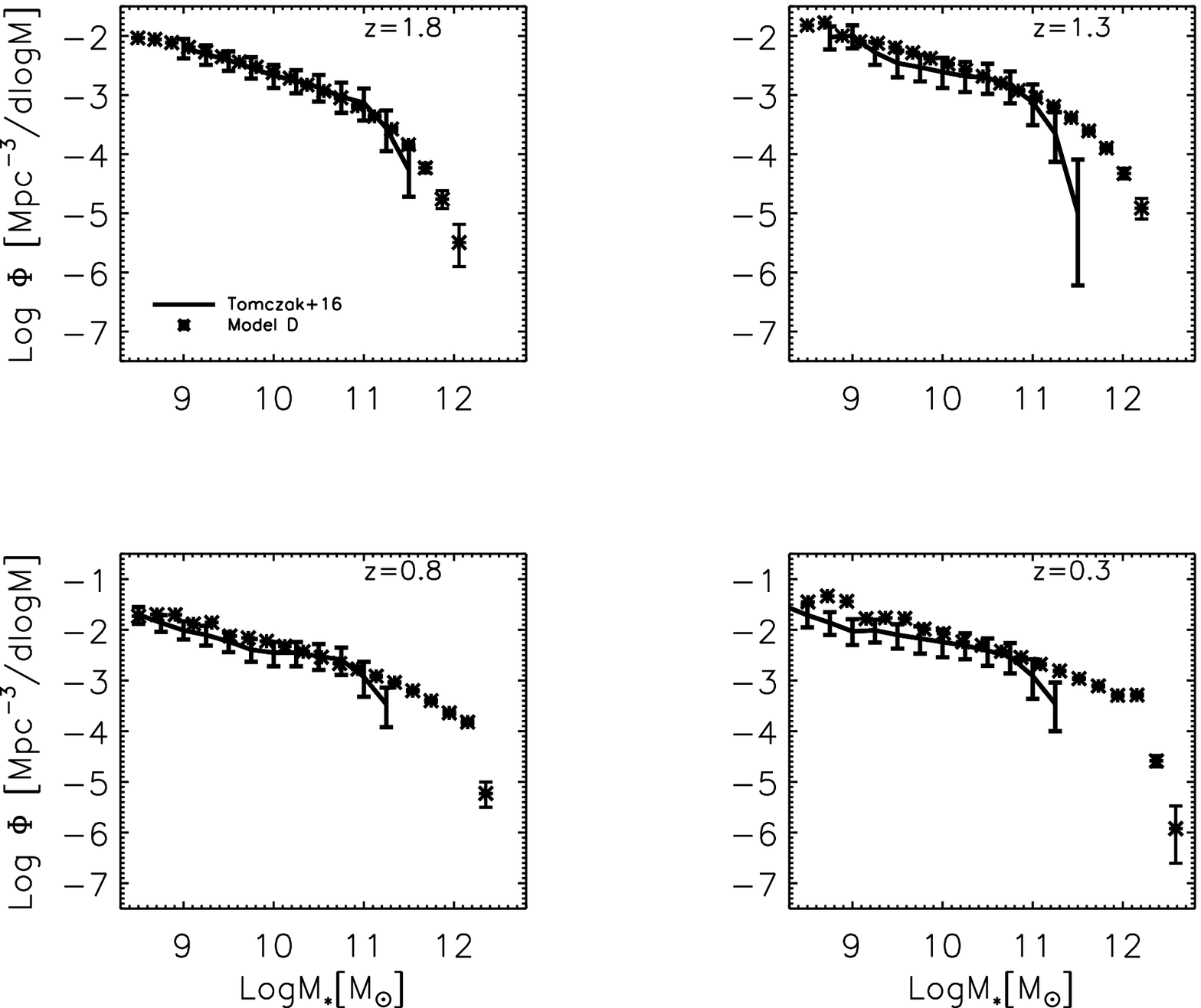} 
\caption{Evolution of the SMF from $z=1.8$ down to $z \sim 0.3$ predicted by our Model B. Here we assume an initial SMF by Tomczak+14 
at $z=2.2$ and coupled to a SFR-$M_*$ relation with a single slope. Stars and solid lines represent model and Tomczak+14 data, respectively.}
\label{fig:tomczak_ss}
\end{center}
\end{figure*}

\begin{figure*} 
\begin{center}
\includegraphics[scale=.6]{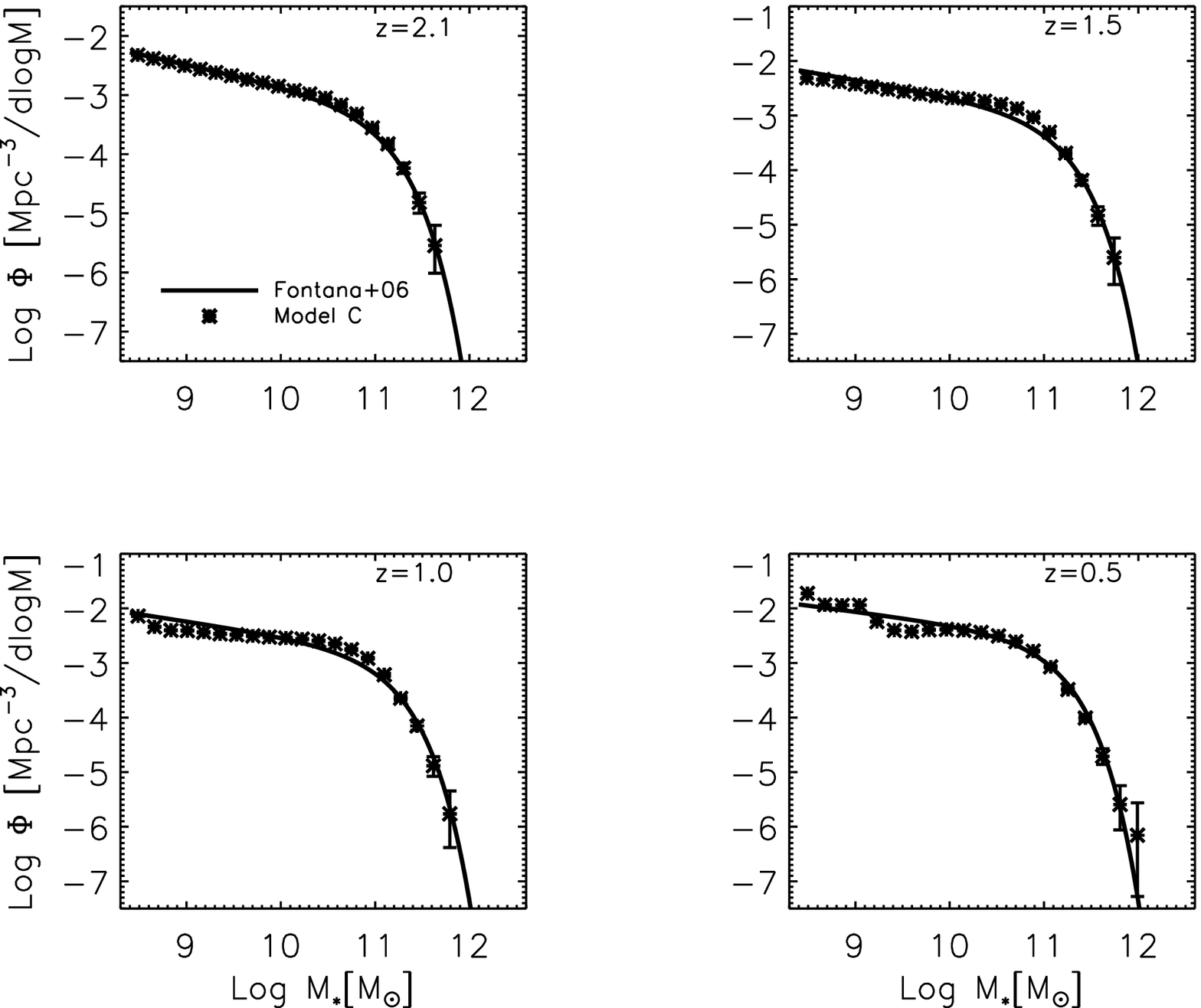} 
\caption{Evolution of the SMF from $z=2.1$ down to $z=0.5$ predicted by our Model C. Here we assume an initial SMF by Fontana+06 
at $z=2.6$ and coupled to a SFR-$M_*$ relation with a mass-dependent slope. Stars represent model data while the solid lines 
represent Fontana+06 fit.}
\label{fig:fontana_mds}
\end{center}
\end{figure*}

\begin{figure*} 
\begin{center}
\includegraphics[scale=.6]{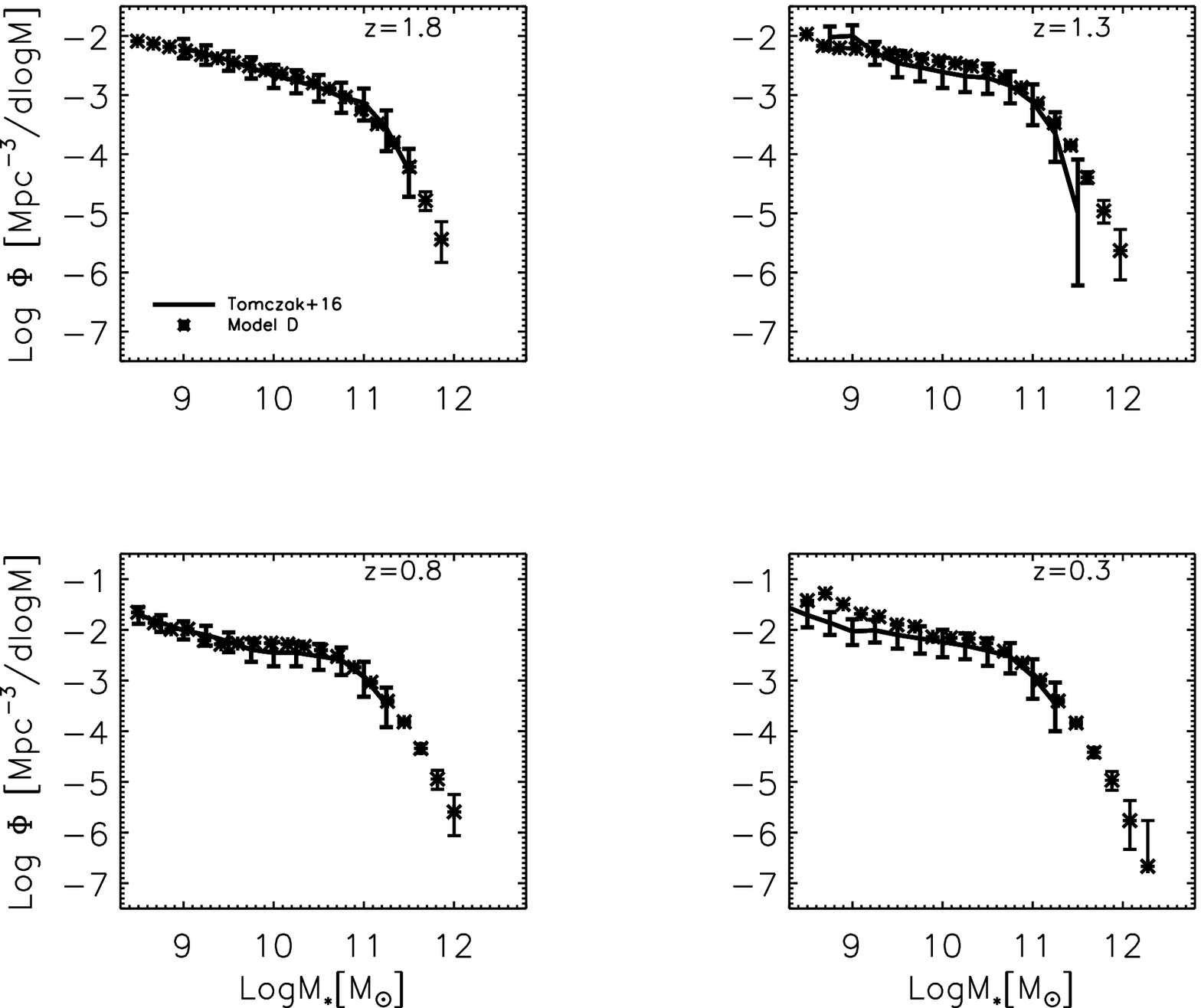} 
\caption{Evolution of the SMF from $z=1.8$ down to $z \sim 0.3$ predicted by our Model D. Here we assume an initial SMF by Tomczak+14 
at $z=2.2$ and coupled to a SFR-$M_*$ relation with a mass-dependent slope. Stars and solid lines represent model and Tomczak+14 data, respectively.}
\label{fig:tomczak_mds}
\end{center}
\end{figure*}

\begin{figure*} 
\begin{center}
\begin{tabular}{cc}
\includegraphics[scale=.48]{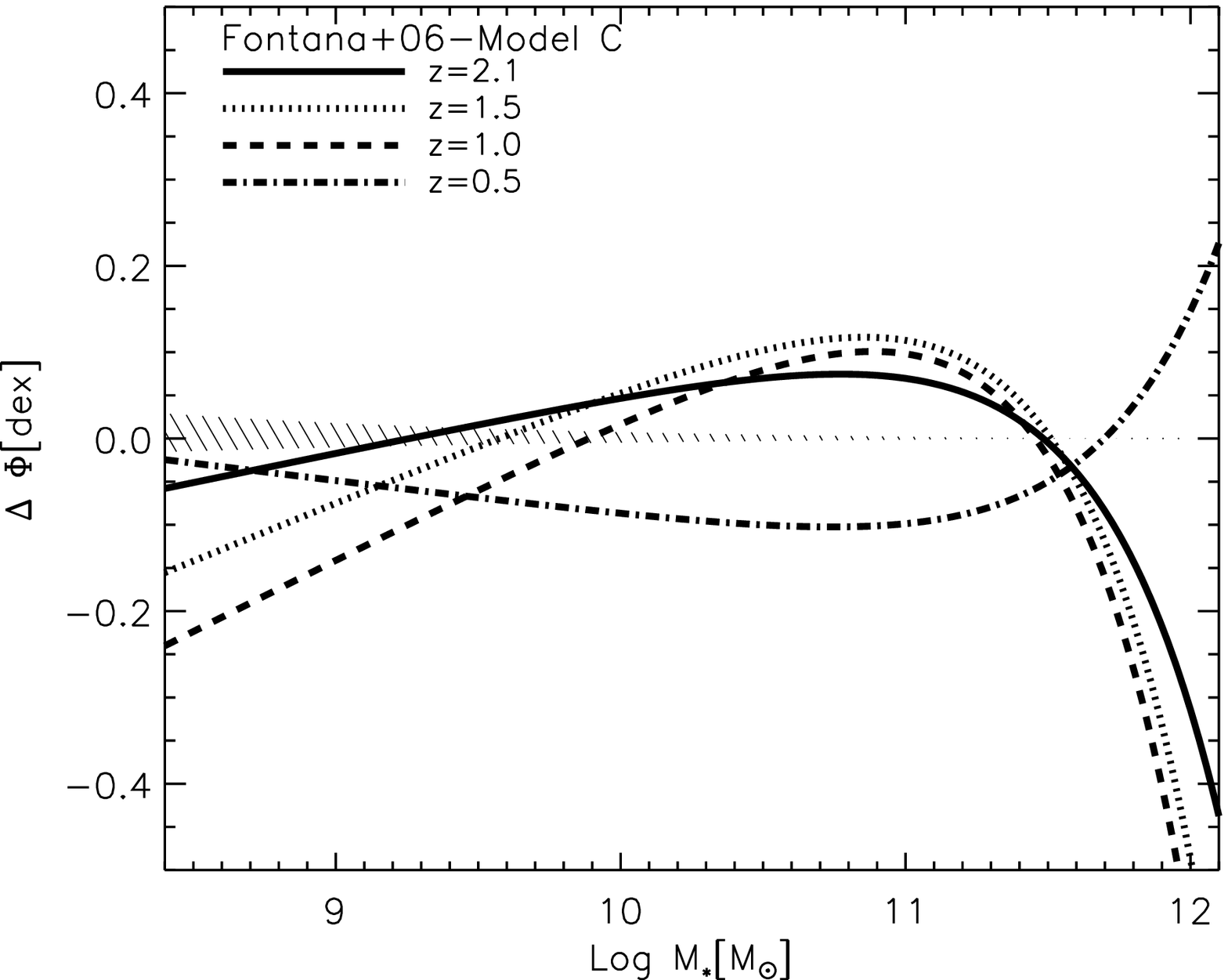} &
\includegraphics[scale=.48]{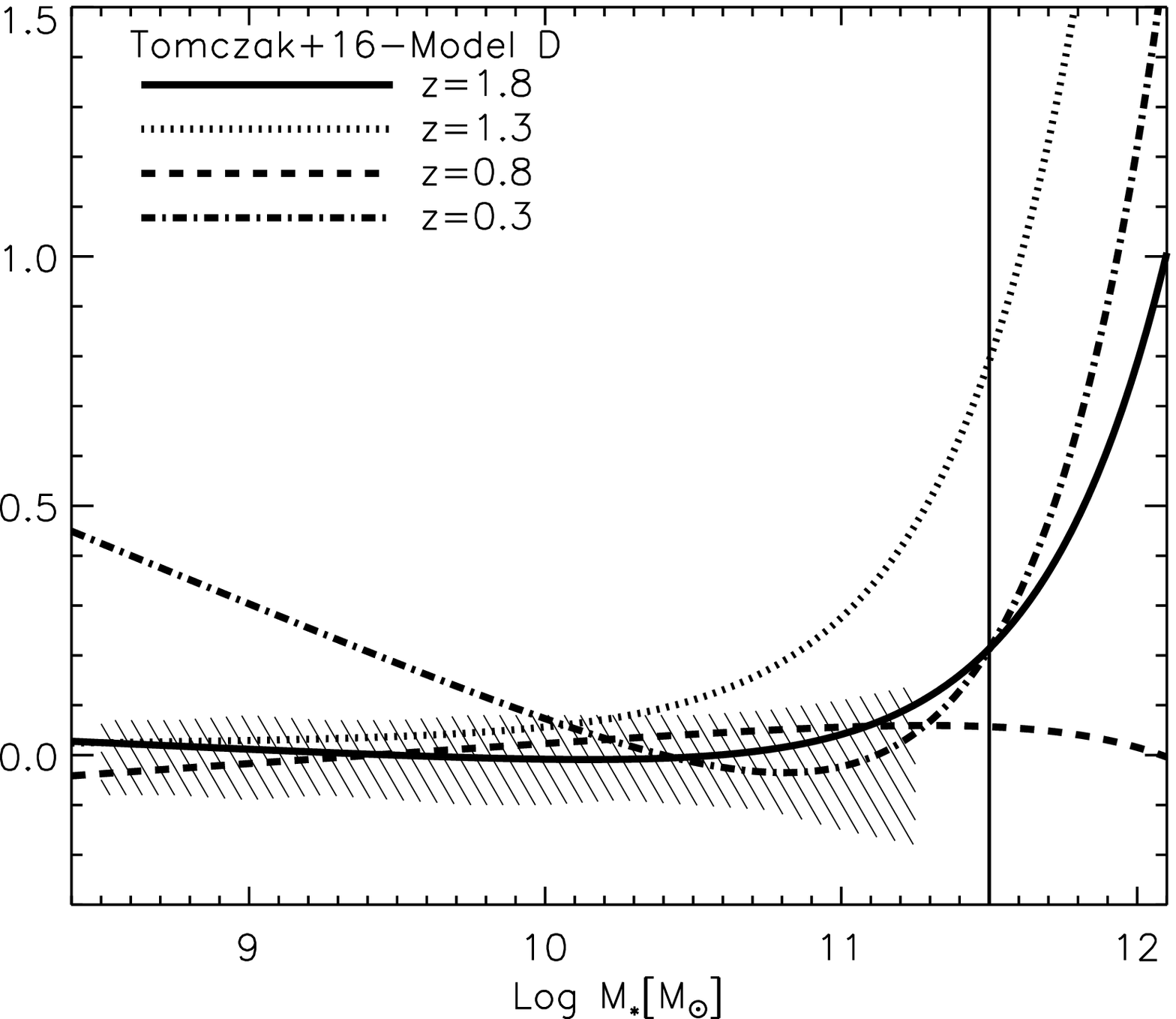} \\ 
\end{tabular}
\caption{Left panel: residuals between the SMF prediceted by our Model C (best-fit Schechter function) and the observed one (best-fit Schechter function), at different redshift (different line 
styles as shown in the legend). Right panel: the same as the left panel, but for Model D. The solid vertical line in the right panel 
indicates the maximum stellar mass bin in Tomczak et al. data, while the shaded region in both panel indicates the scatter around 
observed data at the lowest redshift, $z\sim 0.5$ (left panel) and $z\sim 0.3$ (right panel).}
\label{fig:diff_all_zeta}
\end{center}
\end{figure*}

\begin{figure*} 
\begin{center}
\begin{tabular}{cc}
\hspace{4pt}
\includegraphics[scale=.43]{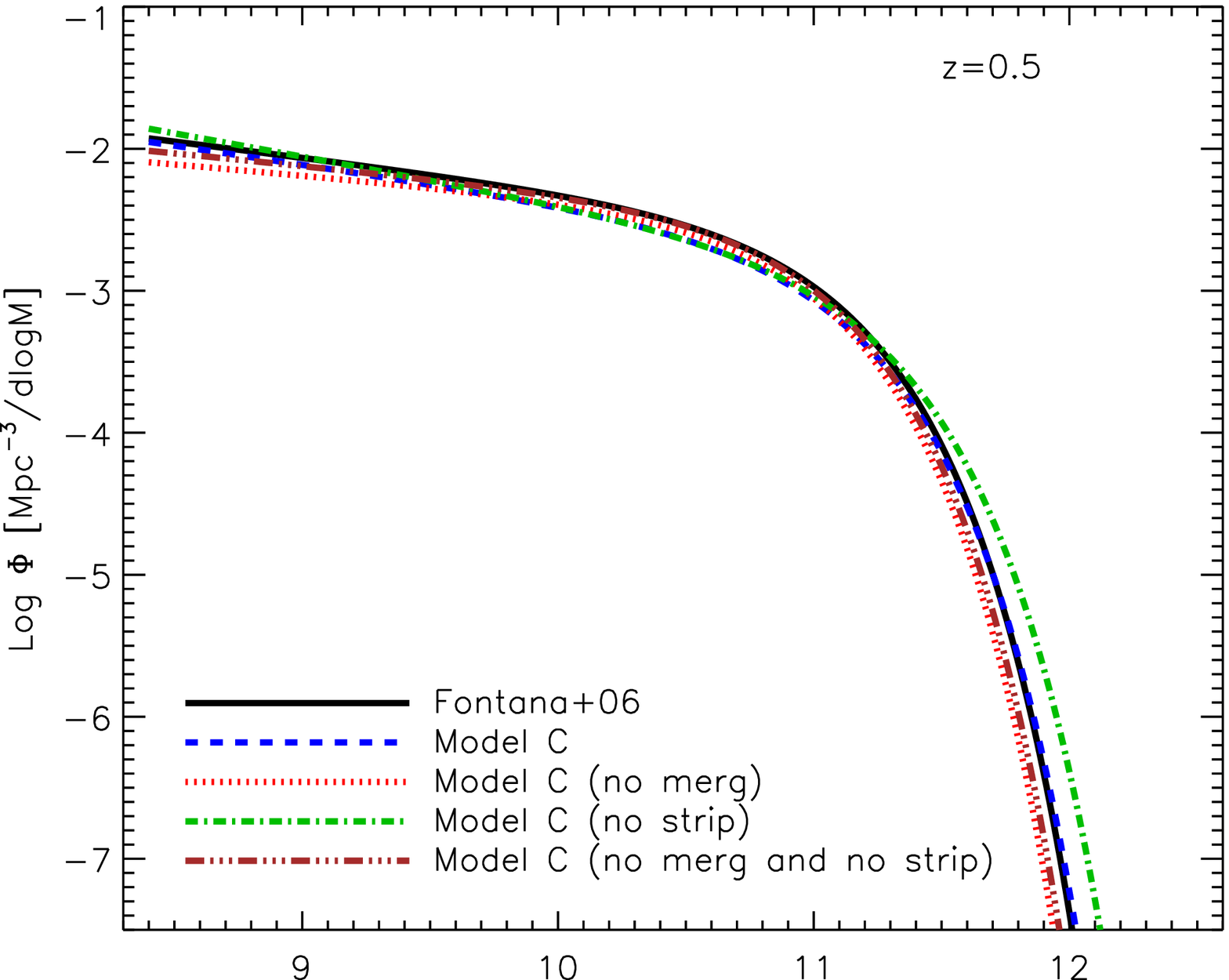} &
\hspace{2pt}
\includegraphics[scale=.43]{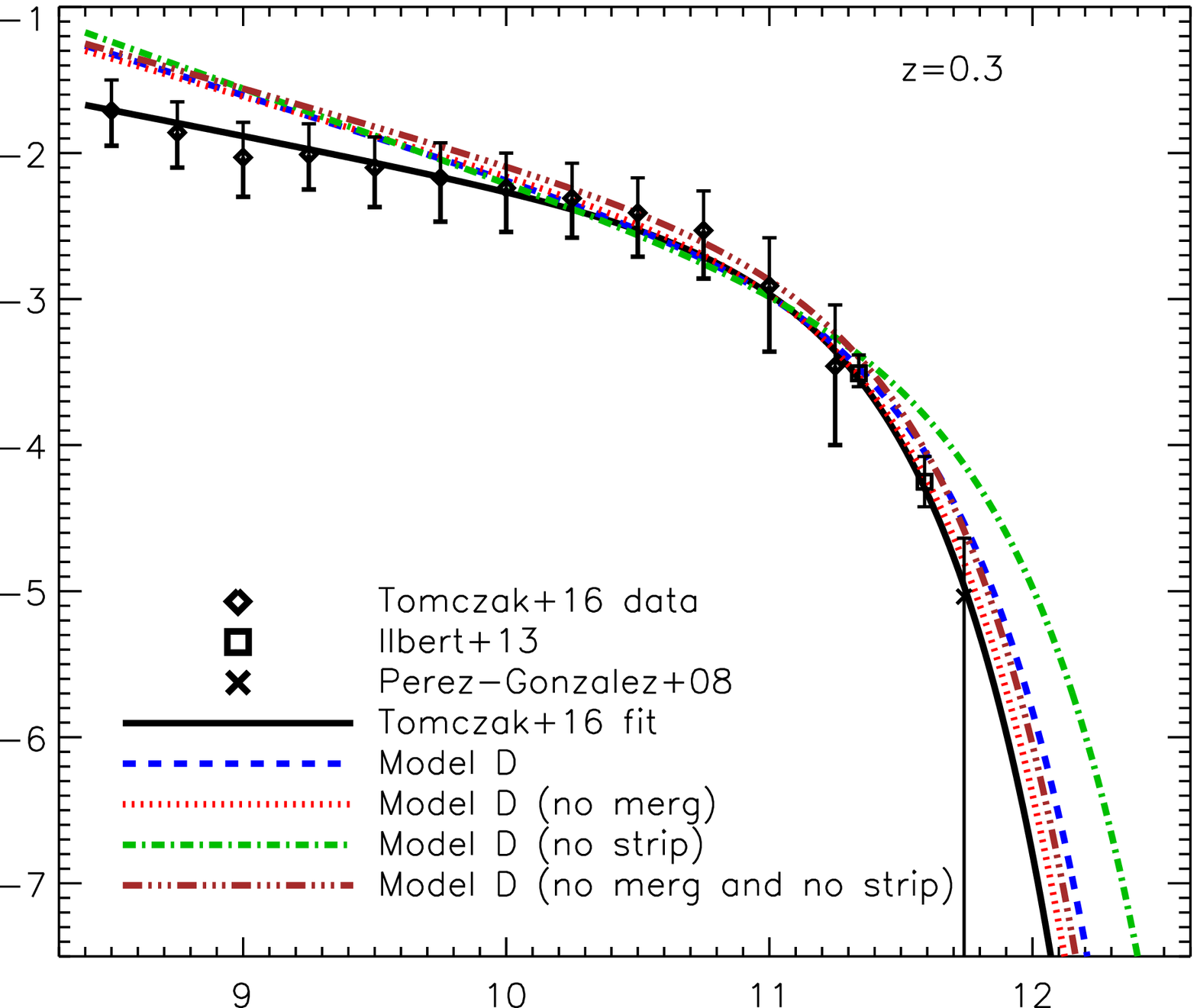} \\ 
\includegraphics[scale=.43]{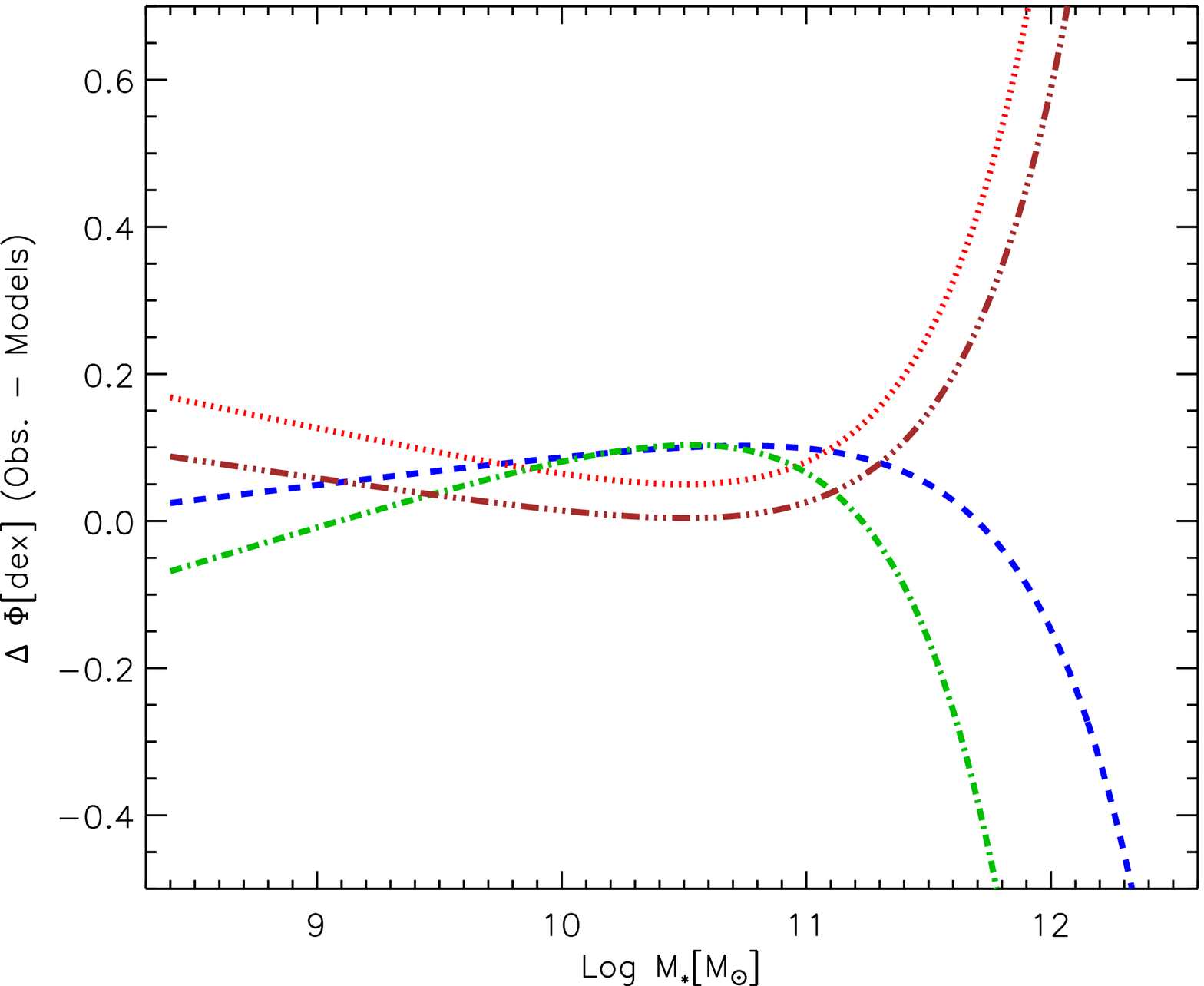} &
\includegraphics[scale=.43]{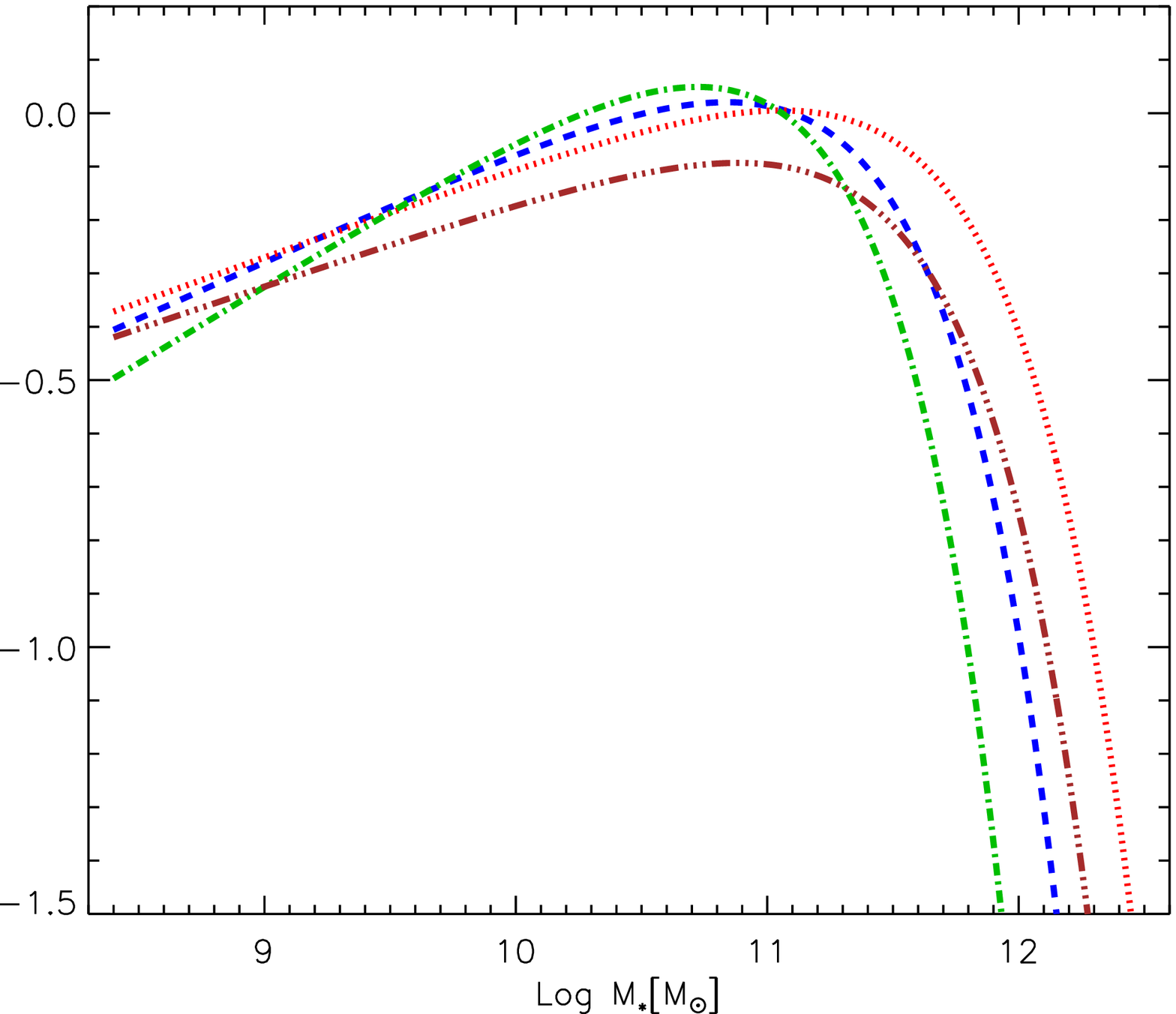} \\ 
\end{tabular}
\caption{Top-left panel: predictions of the SMF at $z=0.5$ by different flavours of Model C (different line styles as shown in the legend), 
compared with the observed SMF. Top-right panel: the same as the left panel but for Model D, that is compared with the observed SMF at 
$z\sim0.3$. Bottom panels: residuals (best-fit Schechter function) between observed data (Fontana+05 on the left and Tomczak+14 on the right) and the different flavours of the
model (different colours).}
\label{fig:models_merg_strip}
\end{center}
\end{figure*}

\begin{figure*} 
\begin{center}
\begin{tabular}{cc}
\includegraphics[scale=.48]{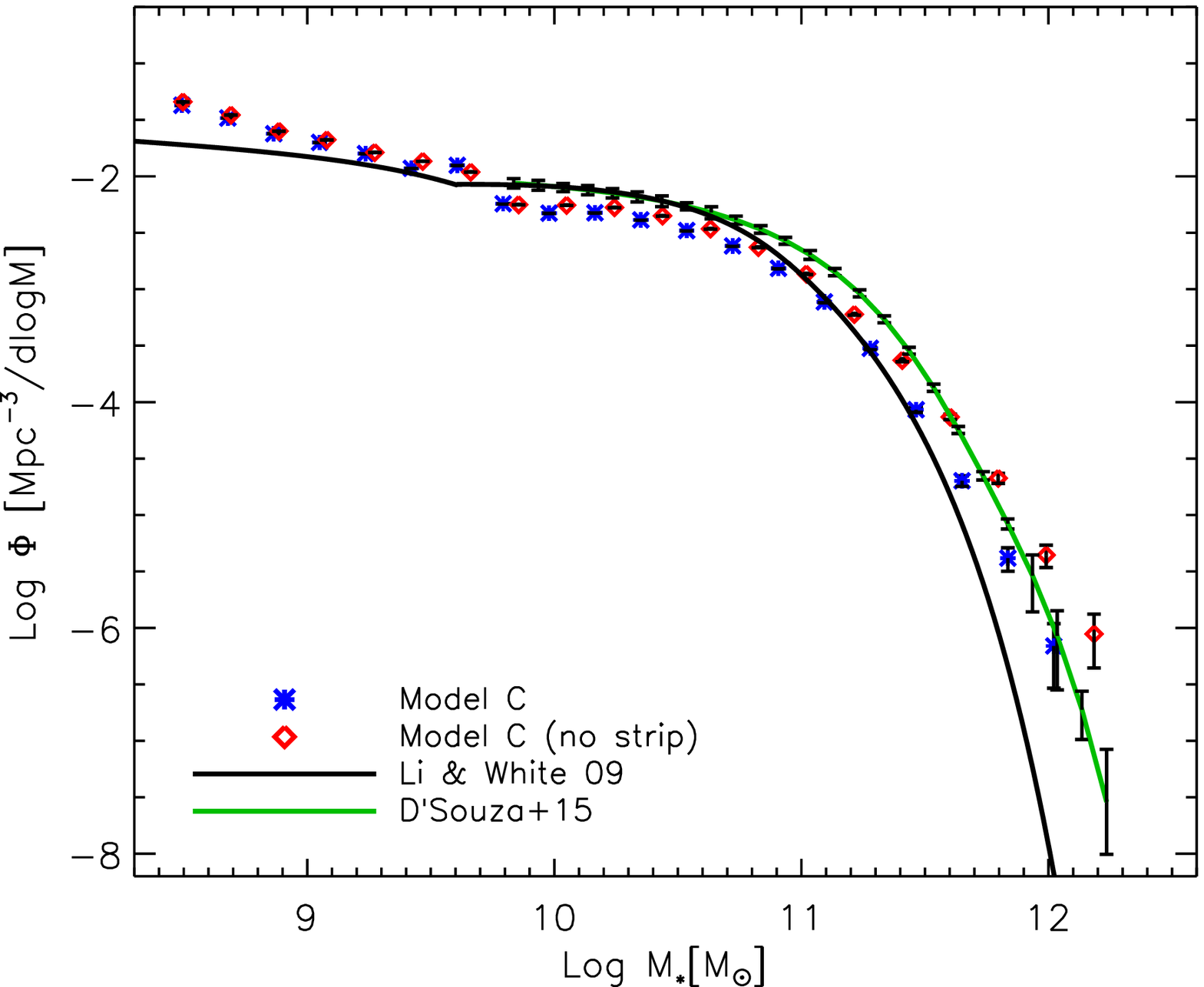} &
\includegraphics[scale=.48]{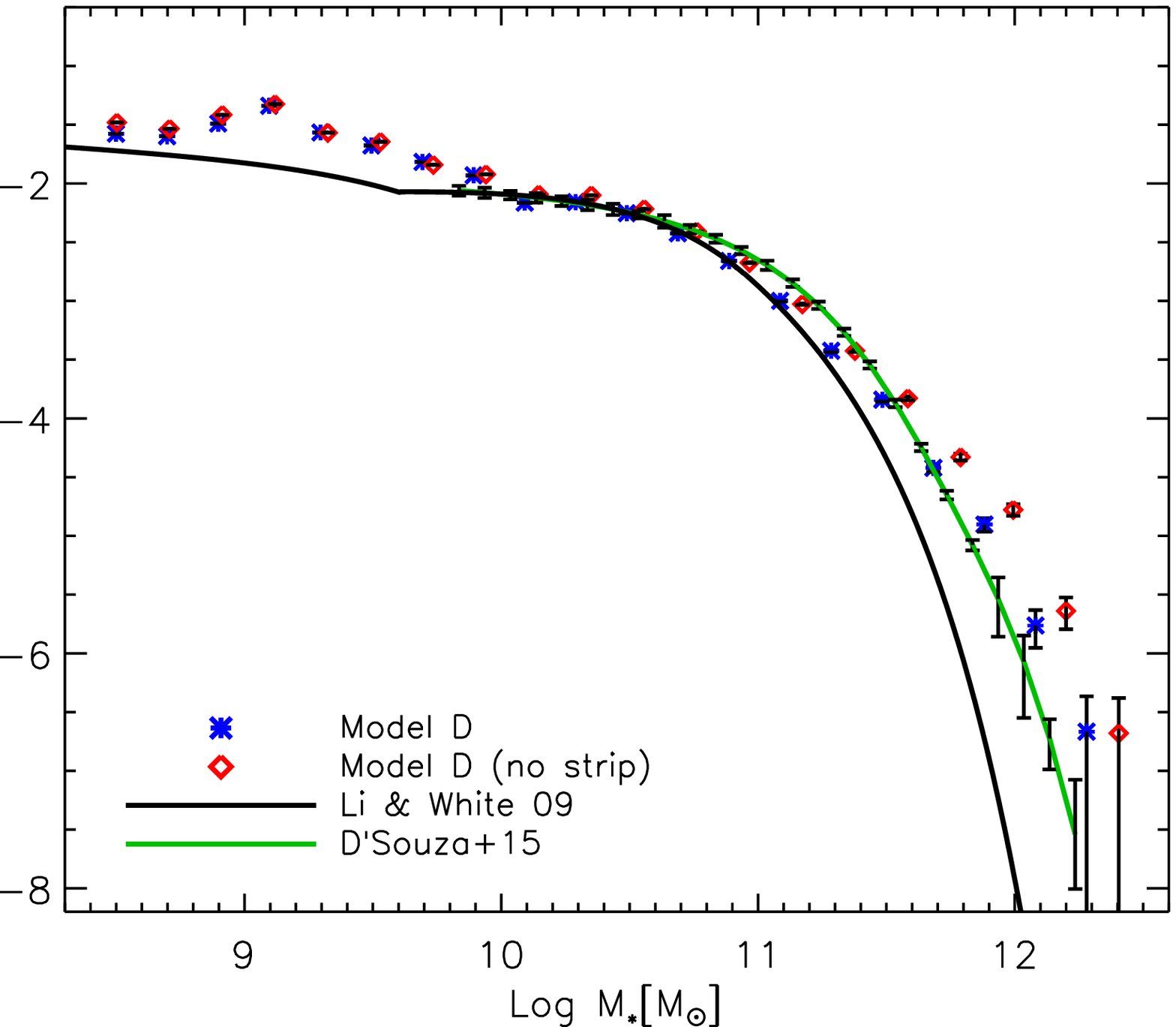} \\ 
\end{tabular}
\caption{Left panel: predictions of the SMF at $z \sim 0.1$ of our Model C with and without stellar stripping (stars and diamonds, respectively), 
compared with the observed SMF by \citealt{li-white09} (solid black line) and \citealt{dsouza15} (solid green line). Right panel: the same as the left panel but 
for Model D.}
\label{fig:z0_predic}
\end{center}
\end{figure*}

\begin{figure*} 
\begin{center}
\begin{tabular}{cc}
\includegraphics[scale=.90]{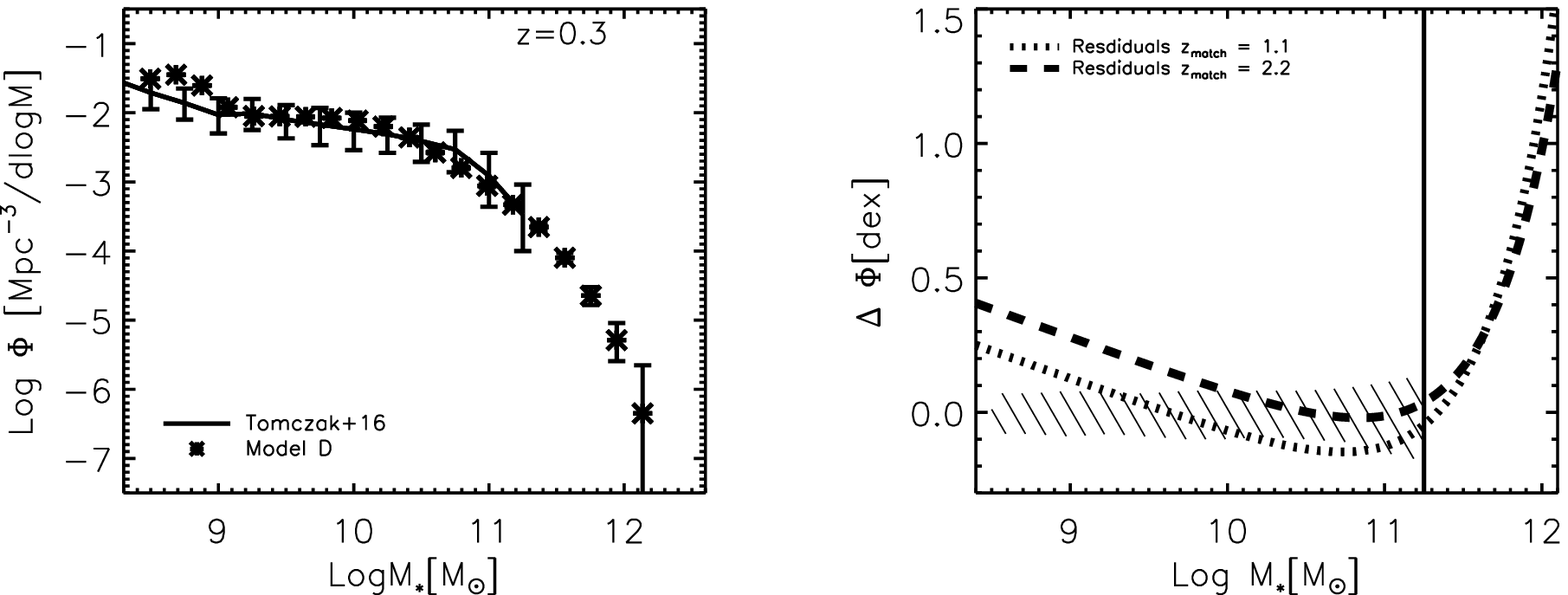} \\
\vspace{0.5cm}
\\
\includegraphics[scale=.70]{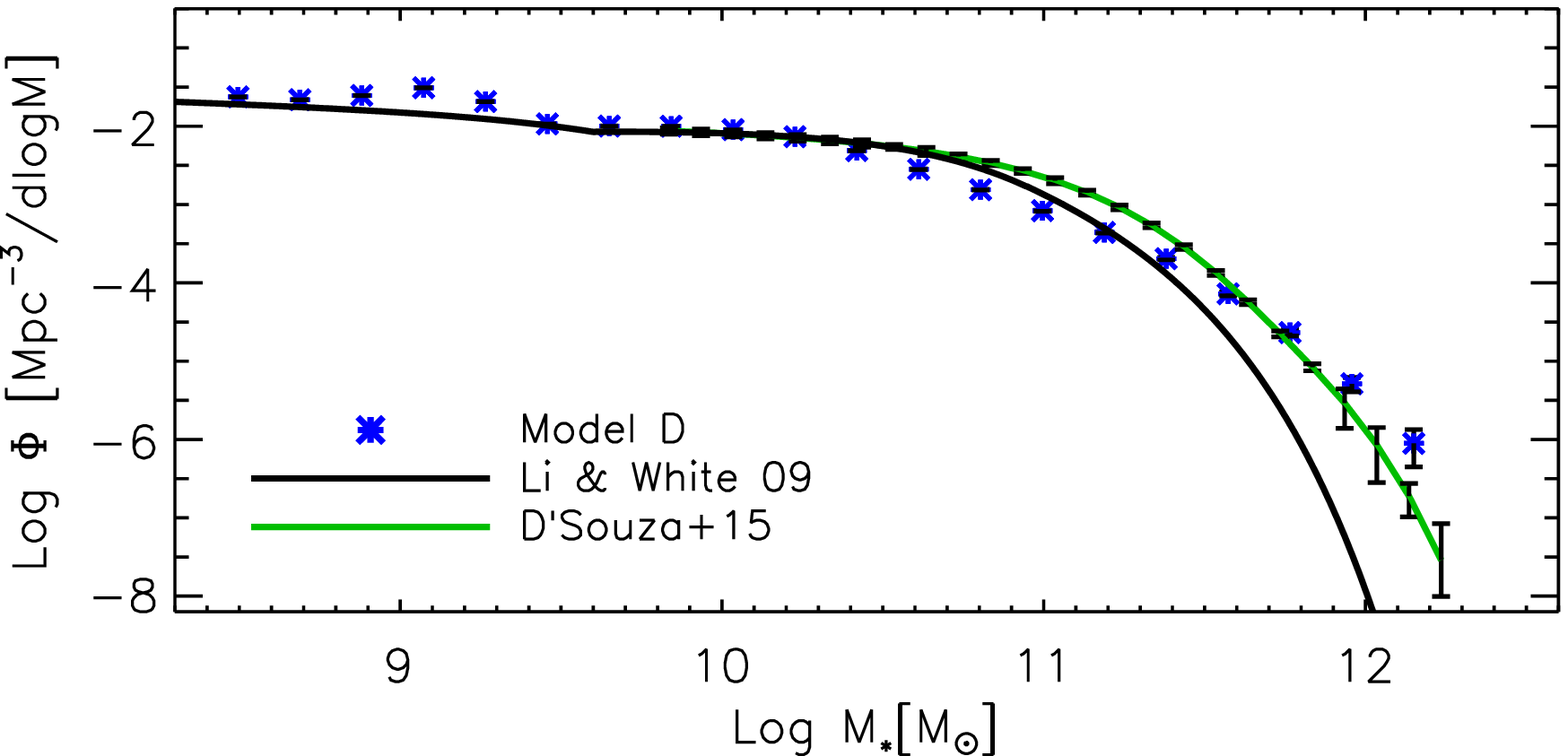} \\
\end{tabular}
\caption{Top-left panel: SMFs at $z \simeq 0.3$. Here we assume an initial SMF by Tomczak+14 coupled to a SFR-$M_*$ relation with a 
mass-dependent slope (Model D), with $z_{match} = 1.1$. Stars and solid lines represent model and Tomczak+14 data, respectively, 
Top-right panel: residuals between the SMF prediceted by our Model D (best-fit Schechter function) and the observed one (best-fit Schechter function) at $z \simeq 0.3$, in case of $z_{match} = 1.1$ (dotted line) and $z_{match} = 2.2$ (dashed line). The solid vertical line indicates the maximum stellar mass bin in Tomczak et al. data at $z \simeq 0.3$, while the shaded region
indicates the observed scatter around data. Bottom panel: SMF at $z=0.1$ as predicted by Model D (stars) and compared with the observed SMF by 
\citealt{li-white09} (solid black line) and \citealt{dsouza15} (solid green line). Here $z_{match}=1.1$.}
\label{fig:lowmatch}
\end{center}
\end{figure*}

\subsection{Role of the SFR-$M_*$ relation}
\label{sec:relation}

In Section \ref{sec:intro} we pointed out that the shape of the SFR-$M_*$ relation in the context of the evolution of the SMF is still under 
debate. A simple power-law (e.g., \citealt{daddi07,elbaz07}) might lead to predictions 
quantitatively different from those obtained by coupling the SMF at high-z to a SFR-$M_*$ relation with a mass-dependent slope (e.g.,
\citealt{lee15,tomczak16}). Here we want to address this issue by looking at the evolution of the SMF as predicted by the four models 
described above.

In Figure \ref{fig:fontana_ss} we plot the evolution of the SMF since $z=2.1$ as predicted by Model A, and compare our results with the 
observed evolution of the SMF by Fontana et al.. Stars represent our model data while the solid lines represent Fontana et al.'s fit.
We start at $z_{match}=2.6$ and, since by construction our model prediction perfectly matches the observed 
SMF at $z=2.6$, we do not show this redshift. As time passes by, our Model A over-predicts the high-mass end of the SMF and becomes unrealistic 
at $z=0.5$. At this time, the SMF looks more like a power-law than a Schechter function. Hence, Model A highly over-predicts the number of massive 
galaxies since $z=1.5$.

Figure \ref{fig:tomczak_ss} shows the evolution of the SMF since $z=1.8$ (and down to $z\sim 0.3)$ as predicted by Model B, and compare our 
results with the observed evolution of the SMF by Tomczak et al. (note that we do not show the panel corresponding to $z=z_{match}$, being
$z_{match} = 2.2$). Stars and solid lines represent model and Tomczak+14 data, respectively.
Over almost all the mass range shown by the observed data points, they agree fairly well with 
our predictions at all redshifts. Nevertheless, Model B shows the same problem as Model A, i.e. the 
number of massive galaxies is excessively over-predicted since $z=1.3$. For both Model A and B we tried to switch off mergers in order not to 
let massive galaxies grow too much, but even in this case predictions do not get significantly closer to observed data (plot not shown). For 
these two models, the bulk of the growth of massive galaxies is due to their intense star formation history.

In Figure \ref{fig:fontana_mds} we plot the evolution of the SMF as predicted by Model C, and similarly to Figure \ref{fig:fontana_ss},
we compare our results with the observed evolution of the SMF by Fontana et al. (line styles and symbols have the same meaning as in Figure 
\ref{fig:fontana_ss}). This model works reasonably well down to $z=0.5$. It matches the high-mass end of the observed SMF at any redshift and shows
small residuals in the low/intermediate stellar mass range. By comparing the evolution of the SMF predicted by Model A (Figure \ref{fig:fontana_ss}) and 
the one predicted by Model C (Figure \ref{fig:fontana_mds}), i.e. the same set of observed SMFs, but a SFR-$M_*$ relation with a single slope (Model A) versus 
a mass-dependent slope (Model C), a non-linear SFR-$M_*$ relation is supported by our results. 

Similarly to Figure \ref{fig:tomczak_ss}, we show in Figure \ref{fig:tomczak_mds} the evolution of the SMF since $z=1.8$ (and down to $z\sim 0.3)$ 
as predicted by Model D, and compare our results with the observed evolution of the SMF by Tomczak et al. (symbols have the same meaning 
as in Figure \ref{fig:tomczak_ss}). Also in this case the high-mass end is better reproduced by the model at all redshifts, and our predictions fairly 
reproduce observations down to redshift $z=0.3$. Overall, Model D confirms that a mass-dependent slope of the SFR-$M_*$ relation, rather than a single slope,
is more consistent with the evolution of the SMF.

Having the idea that Model C and Model D provide an evolution of the SMF more in line with the observed one, we want to quantify in Figure 
\ref{fig:diff_all_zeta} the deviation of the models from the observed SMF as a function of redshift, for Model C (left panel) and 
Model D (right panel). We plot the residuals (best-fit Schechter function) in dex, i.e. the difference between the logarithm of the observed number density and the logarithm 
of the predicted number density, as a function of stellar mass and at different redshifts (as shown in the legend). In the left panel we can see 
that residuals are a function of stellar mass and tend to increase with decreasing redshift (with the exception of $z=0.5)$. The left panel shows that 
Model D has the same trend shown by Model C, but with larger (in modulus) residuals in the high-mass end and lower in the intermediate/low-mass range, 
with respect to Model C (with the exception of the closest redshift to us). By considering the observed scatter (shaded regions) at the lowest redshifts, 
$z\sim 0.5$ (left panel) and $z\sim 0.3$ (right panel), the deviation of Model D from observed data is within the observed scatter in the stellar mass range 
from $10^{10} \, M_{\odot}$ to $\sim 10^{11.2} \, M_{\odot}$, while the deviation of Model C is never within the observed scatter (which is, however, much 
smaller than that of Tomczak et al. data.)

\subsection{Role of Mergers and Stellar Stripping}
\label{sec:merg_strip}

As explained in Section \ref{sec:intro}, most of the previous studies did not consider the role of mergers and stellar stripping in the 
evolution of the SMF. With our approach it is relatively straightforward to isolate their contribution, just by switching on/off the 
prescriptions that account for them. The aim of this section is to prove that their role is not negligible, in particular that of stellar 
stripping.

Figure \ref{fig:models_merg_strip} shows the SMF at $z=0.5$ as predicted by Model C (top-left panel), the SMF at $z \sim 0.3$ as predicted 
by Model D (top-right panel), and the residuals (best-fit Schechter function) between the observed data and the different flavours of the model (bottom panels).
The figure also shows Model C/D when mergers are switched off (dotted lines), with no stellar stripping 
(dash-dotted lines), and when both mergers and stellar stripping are switched off (dash-long-dotted lines). Both panels show that if stripping 
is switched off, the high-mass end of the SMF is over-predicted if compared to the other models. This is easily understandable because satellite galaxies 
do not grow too much, being subject to mass loss. What appears to be very interesting is the roles of mergers and stellar stripping altogether. 
In fact, dashed lines (full model) and dash-long-dotted lines (no mergers and no stripping) lie very close to each other, meaning that 
the two processes have opposite effects: mergers increase the mass of central galaxies (dash-dotted lines), while stellar stripping reduces the 
mass of satellite galaxies. This means that stellar stripping tend to let the SMF move towards the left and mergers towards the right
\footnote{The SMF should move also towards lower number densities because of the smaller number of satellite galaxies. The plot does not
show it since the SMF is dominated by central galaxies, especially in the low-mass end, where the number of satellites is supposed to 
decrease.}. Since they do not have exactly the same effect, quantitatively speaking a full model predicts higher number densities than
a model with no stripping and no mergers, in the high-mass end.

It is worth noting that the intrinsic differences among models are significant only in the high-mass end, where the number density is lower 
and the merging/stellar stripping histories of single galaxies might make the difference. In the rest of the stellar mass range, from the 
low-mass end to the knee of the SMF, models have similar trends and the observed scatter (see the right panel) is not sufficient to rule-out 
one (or more) of them.

In order to understand which model is closer to the observed SMF, we then must focus on the high-mass end. Observed data from ZFOURGE 
at $z \sim 0.3$ (Tomczak et al. data in the right panel of Figure \ref{fig:models_merg_strip}) range from $ \log M_* = 8$ to $ \log M_* = 11.25$. 
Since the upper limit is not sufficient to constrain the very high-mass end, and we need to compare our different predictions with observed data, 
we extend the observed SMF of Tomczak et al. by adding other 3 points (the two squares and the cross in the right panel of Figure \ref{fig:models_merg_strip}) 
which cover the stellar mass range up to  $\log M_* \sim 11.8$.  Squares refer to data from \cite{ilbert13}, UltraVISTA survey,  and the cross refers to the point 
with the largest stellar mass of the SMF by \cite{perez-gonzalez08}, IRAC sample.
These data nicely lie on the fit done by using ZFOURGE data (solid line) and prove that, despite the limited stellar mass range of this set of data, 
the fit is a good representation of the SMF even at high stellar masses. In conclusion, being aware of all the systematic uncertainties in deriving the stellar mass 
(see the discussion in Appendix \ref{sec:append2}), a model with no stripping (dash-dotted line) can be ruled-out and, as a consequence, also a model with 
no stripping and no mergers (dash-long-dotted line).

\subsection{Stellar Mass Function at low redshift}
\label{sec:smf_pt}

The lowest redshifts provided by the two sets of SMFs we have chosen are $z=0.5$ (Fontana et al.) and $z=0.3$ (Tomczak et al.).
Nevertheless, if observations are consistent to each other, it is possible to let the SMF evolve down to the present time and compare the models predictions 
with the observed SMF from other surveys. One important point still under debate is the slope of the high-mass end of the SMF, that appears to be shallower 
than thought earlier. If this is true, this translates in a higher number density of very massive galaxies.

Recently, \cite{dsouza15} compute the SMF at very low redshift for a sample of half a million galaxies from the Sloan Digital Sky Survey (SDSS) in the stellar mass range 
$9.5 < \log (M_* [M_{\odot}h^{-2}) < 12.0$. As the authors point out, systematic differences in the estimation of the stellar mass of a galaxy can be
due to a different choice of the IMF (for which there are, however, well constrained corrections), to the stellar mass-to-light ratio (M/L), and to 
different estimations of the galaxy total flux (see also discussion in Appendix \ref{sec:append2}). The key point of their work is the latter. Therefore, for the sample of galaxies used by \citet{li-white09}, 
they derive flux corrections to the model magnitudes by stacking together mosaics of similar galaxies in bins of stellar mass and concentration, and re-derive 
the galaxy stellar mass function at redshift $z=0.1$. They find that the flux corrections result in a higher massive end of the SMF and make the slope shallower 
than that found by \citealt{li-white09}, but steeper than that estimated by \cite{bernardi13}.

In Figure \ref{fig:z0_predic} we plot our models predictions at redshift $z \sim 0.1$ for Model C (left panel) and Model D (right panel), with (stars) and 
without (diamonds) stellar stripping, and compare our results with the SMF computed by both \citet{li-white09} (solid line) and \citet{dsouza15} (triangles). 
As we highlighted above, if stellar stripping is switched off it would result in a higher massive end, but too high if compared with the observed one, for either
models, C and D. A model in which stellar stripping is switched on, instead, agrees fairly well with the observed SMF by \citet{dsouza15} in the intermediate 
stellar mass range (as a model with no stripping does), and lies closer to data than a model with no stripping up to the very massive end. It must be noted that 
the best prediction is given by Model D, while Model C under-predicts the number density in the intermediate stellar mass range. Therefore, according to these 
results, our models predictions, especially Model D, support an higher massive end, more in agreement with the observed high-mass end by \citet{dsouza15} than 
that by \citet{li-white09} or by \citet{bernardi13}. In principle, a model with no stellar stripping would be closer to the SMF by \citet{bernardi13} in the high
mass end, being it higher than that found by \citet{dsouza15} (see Figure 7 of D'Souza et al.). Nevertheless, such a model cannot account for the abundance of 
satellite galaxies in haloes of different mass (see Figure \ref{fig:CSMF} in Appendix \ref{sec:append}) that we have used to calibrate the stripping efficiency 
in our model of stellar stripping.

Problems arise in the low-mass end, where both models (C and D) over-predict the abundance of dwarf galaxies. This problem is more serious for Model D, for which 
the number density of galaxies with mass $\log M_* < 10$ is over-predicted by up to a factor 0.5 dex (at $\log M_* \sim 9.1$). We will come back on this issue in 
the next section.

\section{Discussion}
\label{sec:discussion}

In order for the SFR-$M_*$ relation to be consistent with the observed evolution of the SMF, it would be important to consider all the uncertainties 
in measuring both the stellar mass of galaxies and their star formation rate. This point has been discussed by \cite{leja15}, and somewhat 
considered by \cite{tomczak16}. Given all the systematic uncertainties, it is possible, however, to discriminate between a SFR-$M_*$ relation with  
a single power-law shape, and one with a mass-dependent slope. This is one of the main goals of this paper. 

In Figures \ref{fig:fontana_ss} and \ref{fig:tomczak_ss} we have shown that a single power-law shape cannot describe the evolution of the SMF, since 
for both sets of SMFs chosen this would bring to highly over-predict the high-mass end, giving to the SMF an unrealistic shape at low redshift.
Nevertheless, these figures show that, especially with the set of SMFs by \cite{tomczak16}, the slope of the SFR-$M_*$ relation (0.9) for low-mass 
galaxies seems to be acceptable, since the low-mass end of the SMF is reproduced at least down to $z \sim 0.8$. Likely, a slightly lower 
normalisation would be better because predictions lie lightly above observations at those redshifts and mass ranges (this is clearly shown in the case 
of our Model B, when the set of SMFs by Tomczak et al. is coupled to a single power-law). The actual problem does not concern the choice of the slope 
of the SFR-$M_*$ relation itself, rather its whole shape and evolution with redshift. 

To shed light on this issue, we have coupled the observed SMF to a SFR-$M_*$ relation with a mass-dependent slope (Model C/D). Figures \ref{fig:fontana_mds}
and \ref{fig:tomczak_mds} show that such a relation gives much better predictions. First, the overall shape of the predicted SMF is comparable with the observed 
one, and second, the high-mass end at every redshift is better reproduced. Despite that, it is not enough to make the SFR-$M_*$ relation and the observed 
evolution of the SMF be consistent one another. This has been shown in Figure \ref{fig:diff_all_zeta}, where the residuals between the observed and predicted 
SMFs increase with decreasing redshift, although we consider galaxy mergers and stellar stripping. Tomczak et al, who find a similar mismatch at all redshifts,
argue that this disagreement implies that either the SFRs are overestimated and/or the observed growth of the \cite{tomczak14} SMF is too slow. Similar arguments 
have been discussed by other authors. \cite{weinmann12} show that, in order to reconcile the SFR-$M_*$ relation with the growth of the SMF at $z<1$, either 
the slope of this relation is greater than 0.9, or a high rate of destruction by mergers must be invoked. We have shown in Figure \ref{fig:models_merg_strip} 
that mergers cannot have such a relevant role, and the match substantially improves in the high-mass end when stellar stripping is considered. 

Similarly to Tomczak et al., we believe that the discrepancies are mainly due to errors in stellar mass and SFRs estimates. Errors in the stellar mass estimate 
reduce the accuracy of the observed SMF, while errors in the SFR estimate will collect additional scatter around the SMF during its evolution.
The observed SMF at high-z is not completely constrained, and this is proved by the fact that different observed SMFs do not match each other, as those we have chosen.
Tomczak et al. find similar residuals, in particular in the low/intermediate stellar mass range. They use the same evolving SFR-$M_*$ relation with redshift 
(Equation \ref{eq:dep_slope}) to generate star-fomation histories of galaxies, and integrate the set of star-formation histories with time. They then obtain 
mass-growth histories to compare with the mass growth from the evolution of the stellar mass function of \cite{tomczak14}. This method is in spirit similar to that 
used in this paper, but they let the SMF evolve for a limited time (see their Figure 10). Moreover, their method does not account for galaxy mergers and stripping. 
By comparing the observed and inferred SMFs, they conclude that a reasonable match would require between 25-65 per cent of the excess galaxies to merge with a more 
massive galaxy per Gyr. This definitely exceeds current estimates of galaxy merger rates (e.g. \citealt{lotz11,williams11,leja15}).

In our study we let the SMF evolve with time starting from high-redshift and down to low-redshift, according to the same evolution with redshift given by the 
SFR-$M_*$ relation used by Tomczak et al.. In the case of Model D, that is the set of SMFs by \citealt{tomczak14} coupled to the SFR-$M_*$ relation given by 
Equation \ref{eq:dep_slope}, we start at $z_{match} \simeq 2.2$ and let the SMF grow down to $z \simeq 0.3$. As said above, during this time systematic uncertainties 
both in the stellar mass and SFR estimates can propagate, thus increasing the mismatch between the observed and inferred SMFs. If we restrict the evolution of the 
SMF by lowering $z_{match}$, where both the stellar mass and SFR measurements should be more reliable, we then expect a better match, which translates in smaller 
residuals. We test this argument in Figure \ref{fig:lowmatch}, where $z_{match}$ has been set to 1.1, and the SMF evolves down to $z\simeq 0.1$, for Model D. 
As we can see, the SMF at $z=0.3$ predicted by the model (top-left panel) is more in agreement with the observed one, being residuals around 0.2 dex smaller 
(top-right panel) and within the observed scatter in a larger stellar mass range than before. This implies that if we let the SMF evolve for a shorter time, 
errors on the measurements of the SFRs have a shorter time for propagating, resulting 
in smaller residuals. Another likely explanation is that there might be discrepancy between stellar mass and SFR at high-z, but not for $z<1$. This conclusion
has been found in other works (see, e.g., \citealt{leja15,tomczak16} and references therein). From low stellar masses to $\log M_* = 11.25$, which corresponds to the maximum stellar mass bin in 
Tomczak et al. data at $z \simeq 0.3$ (vertical solid line in the top-right panel of Figure \ref{fig:lowmatch}), and where both fits are more accurate, residuals decrease from 
a maximum of 0.4 dex (dashed line, low-mass) to 0.25 dex (dotted line, low-mass), while the minimum is close to 0 in both cases (at $\log M_* \sim 10.8$). It is worth 
noting also that we obtain comparable (with respect to Tomczak et al.) residuals if $z_{match}=2.2$, and smaller if $z_{match}$ is lower, because we are taking into account 
most of the processes that can influence galaxy growth, such as mergers and stellar stripping (not considered by Tomczak et al.).

In the bottom panel of Figure \ref{fig:lowmatch} we plot the SMF at $z=0.1$ as predicted by Model D (with $z_{match}=1.1$), and compared with the observed ones 
(by \citealt{li-white09} and \citealt{dsouza15}), as done in Figure \ref{fig:z0_predic}. In Section \ref{sec:smf_pt} we highlighted that Model D over-predicts the 
low-mass end of the SMF at $z=0.1$ by a non-negligible factor (around 0.5 dex at $\log M_* \sim 9.1$). This panel shows that, when the match is done at a lower redshift,
the low-mass end is more in agreement with observed data. Overall, Figure \ref{fig:lowmatch} demonstrates that our modelling is very 
sensitive to the shape of the  SMF and in particular the slope of the low-mass end, not only to the SFR-$M_*$ relation. The slope of the low-mass end constrains the number of 
low-mass galaxies that have to grow. A steep slope of the SMF at high-redshift would bring to over-predict the number density of low/intermediate stellar mass galaxies, 
and a shallow slope would act in the other direction. 

\section{Conclusions}
\label{sec:conclusions}

We have analysed the stellar mass growth of galaxies from $z>2$ taking advantage of the Subhalo Abundance Matching method.
We have linked the observed SFR-$M_*$ relation to the observed SMF at high-redshift to study its evolution with time. In
this paper, two sets of SMFs and two different SFR-$M_*$ relations have been chosen: the set of SMFs by \citealt{fontana06} (GOODS-MUSIC 
catalog), and the set of SMFs by \citealt{tomczak14} (ZFOURGE catalog). Both sets of SMFs have been coupled to: (1) a SFR-$M_*$ relation 
with a single power-law shape, and a redshift-dependent normalisation (Equation \ref{eq:single_slope}), and: (2) SFR-$M_*$ relation 
redshift-dependent and with a mass-dependent slope (Equations \ref{eq:dep_slope} and \ref{eq:parameterisation_all}).
From our analysis we conclude that:
\begin{itemize}
 \item The evolution of the stellar mass function is much more consistent with a SFR-$M_*$ relation redshift-dependent and with a 
 mass-dependent slope, rather than a simple single power-law. If coupled to a SFR-$M_*$ with a single power-law, the SMF would not have 
 a Schechter shape at low-redshift. Moreover, the number density of high stellar mass galaxies ($\log M_* > 11.2-11.3$) would be greatly 
 over-predicted at any redshift. This is in agreement with predictions of other models of galaxy evolution, semi-analytic, hydrodynamical, and 
 abundance-matching models (e.g. \citealt{weinmann12,leja15}), and supported by observations (e.g \citealt{whitaker14,tomczak16}).
 
 \item Galaxy mergers and stellar stripping are physical processes that must be taken into account in order to fully analyse the evolution 
 of the SMF. We have shown that both mergers and stripping are important in shaping the massive end of the SMF. 
 
 \item The observed SMF at high-z is not 100 per cent accurate. We have tested two different sets of SMFs coupled to the same SFR-$M_*$ relation 
 and same modelling for mergers and stellar stripping. They result in different evolutions down to low-redshift.
 
 \item The observed high-mass end at very low redshift is not yet accurate. We have shown that our models predictions favour an higher massive 
 end than that estimated by \citealt{li-white09}, more in agreement with recent results by \citealt{dsouza15}.
 
 \item The inferred evolution of the stellar mass function is sensitive to the shape of the observed SMF 
 at $z=z_{match}$. Moreover, during the evolution of the SMF systematic uncertainties both in the stellar mass and SFR estimates can propagate, 
 thus increasing the mismatch between the observed and inferred SMFs. We have shown that, if the inferred SMF starts to evolve when stellar mass 
 and SFR measurements are more reliable, and uncertainties have less time to propagate, the match at low-redshift is characterised by rather 
 smaller residuals.
\end{itemize}

Although the match between the observed and inferred SMFs has improved with respect to previous studies, given by the fact that mergers are taken 
into account by using accurate merger trees and stellar stripping is modelled, there is still something missing. It might be due to one or more of 
the aforementioned problematics, or to any other process that we are not yet considering. In a forthcoming paper, we will re-address this issue by means 
of a slightly different approach. We use the set of SMFs by \citealt{tomczak14} coupled to the observed SFR-$M_*$ with a mass-dependent slope. 
In this way we will initialise our sample of model galaxies at $z=z_{match}$. Making use of some parameterisations of the star formation histories for 
satellite and central galaxies (such as those suggested by \citealt{wang07} or by \citealt{yang12}), we will use our model to understand 
when star formation has to be stopped and for which galaxies, forcing the model itself to match the SMF at low-redshift. We will then split the sample 
of galaxies in star-forming and passive according to some cuts in colour, with the purpose to match the observed SMFs and the galaxy two-point correlation 
functions of both types of galaxies.

\section*{Acknowledgements}
EC acknowledges financial support by Chinese Academy of Sciences Presidents' International 
Fellowship Initiative, Grant No.2015PM054.
EC, XK and AR acknowledge financial support by 973 program (No. 2015CB857003, 2013CB834900),
NSF of Jiangsu Province (No. BK20140050), the NSFC (No. 11333008,11550110182,11550110183), and
the Strategic Priority Research Program the emergence of cosmological structures€ of the 
CAS (No. XDB09010403).

\label{lastpage}

\appendix

\section{Satellite Abundance in Groups and Clusters}
\label{sec:append}

\begin{figure*}
\begin{center}
\begin{tabular}{cc}
\includegraphics[scale=.42]{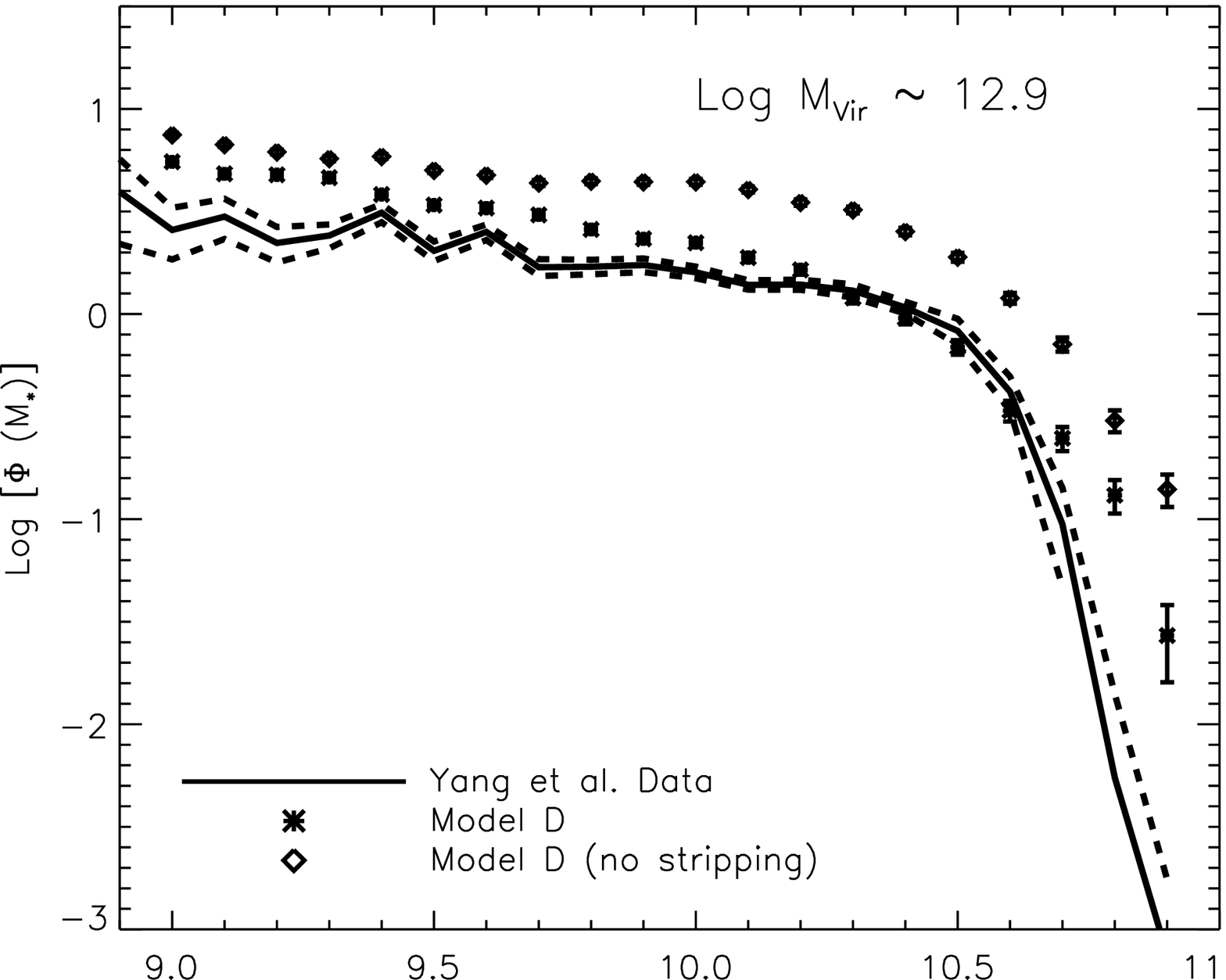} &
\includegraphics[scale=.42]{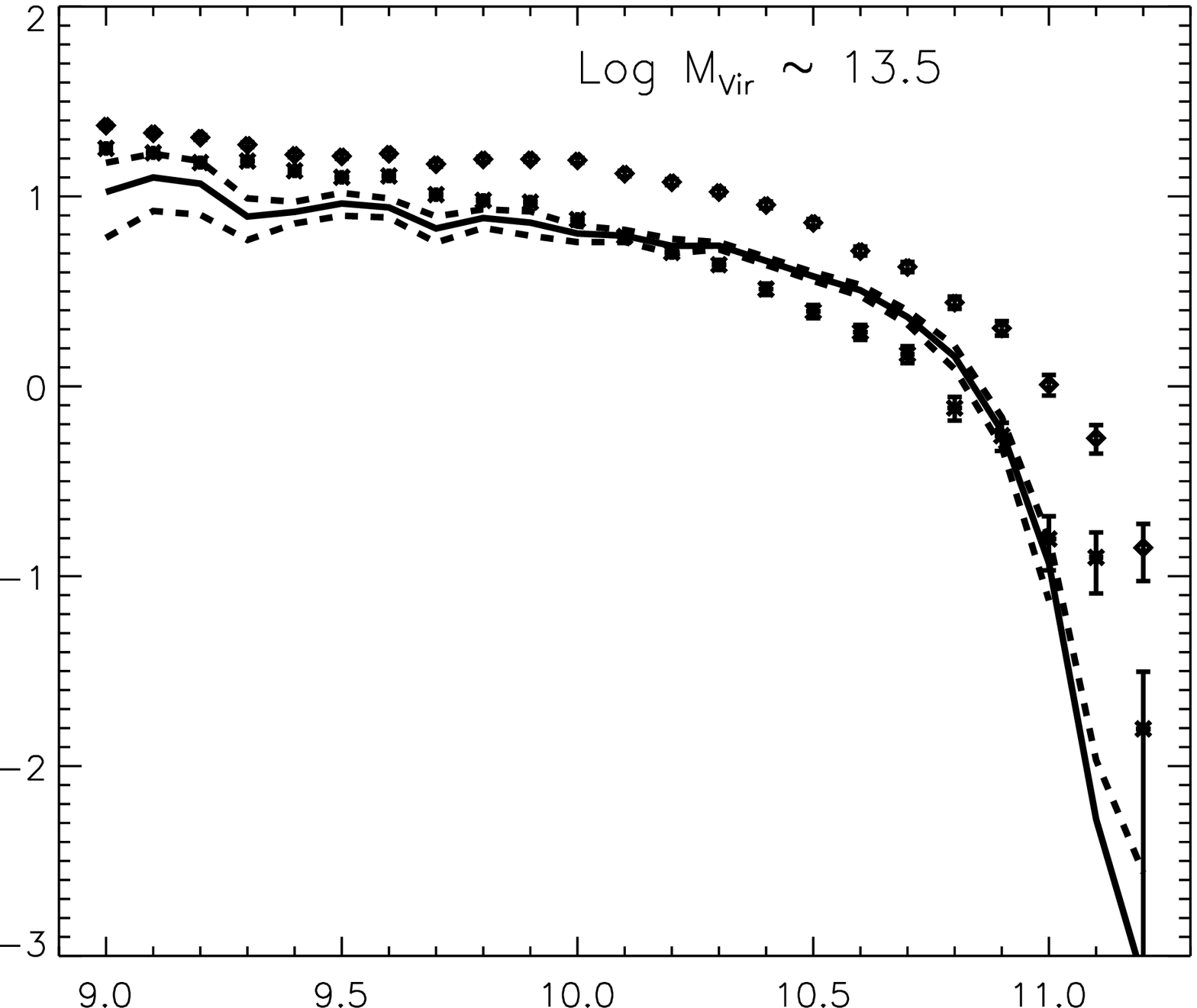} \\
\includegraphics[scale=.42]{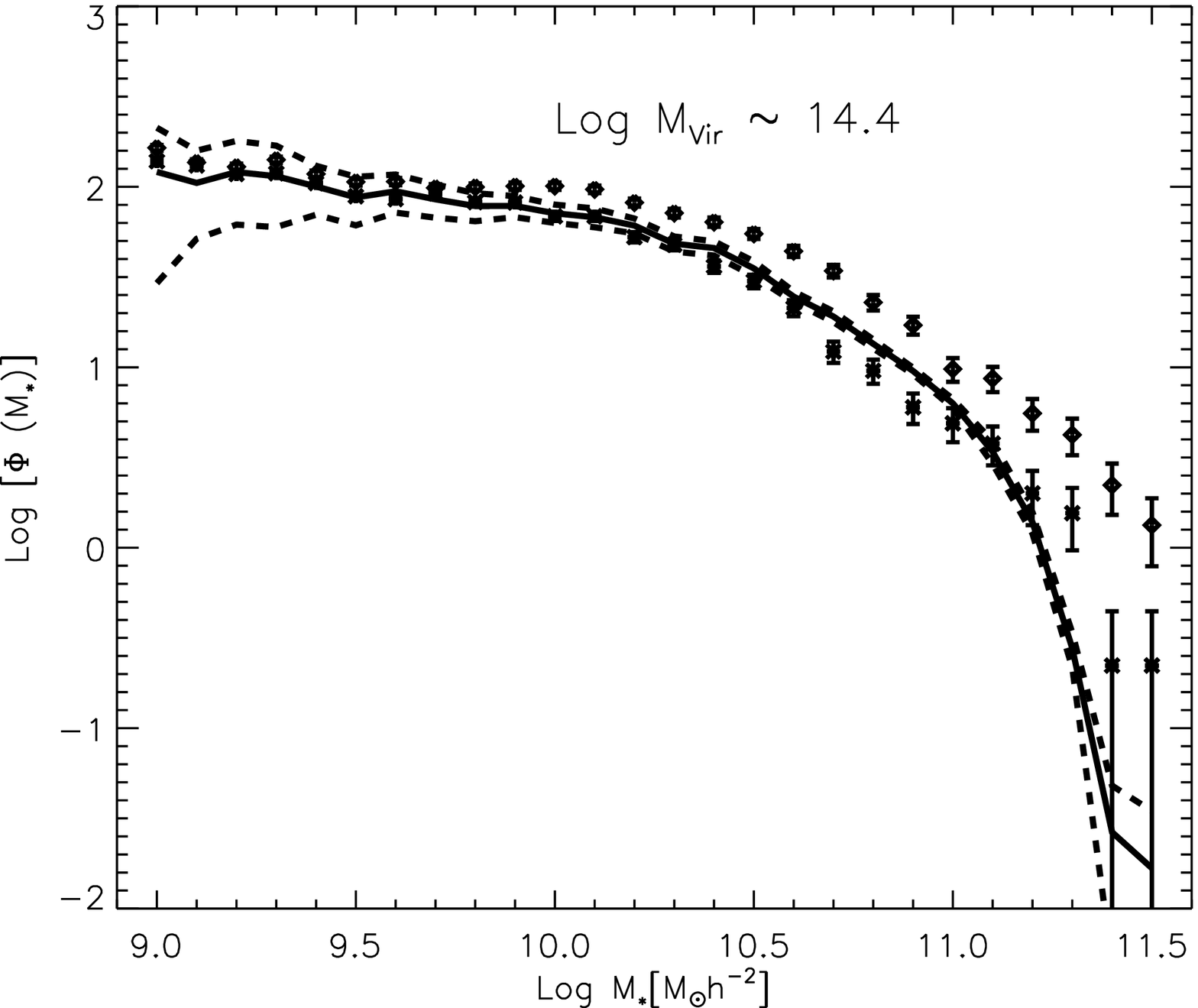} &
\includegraphics[scale=.42]{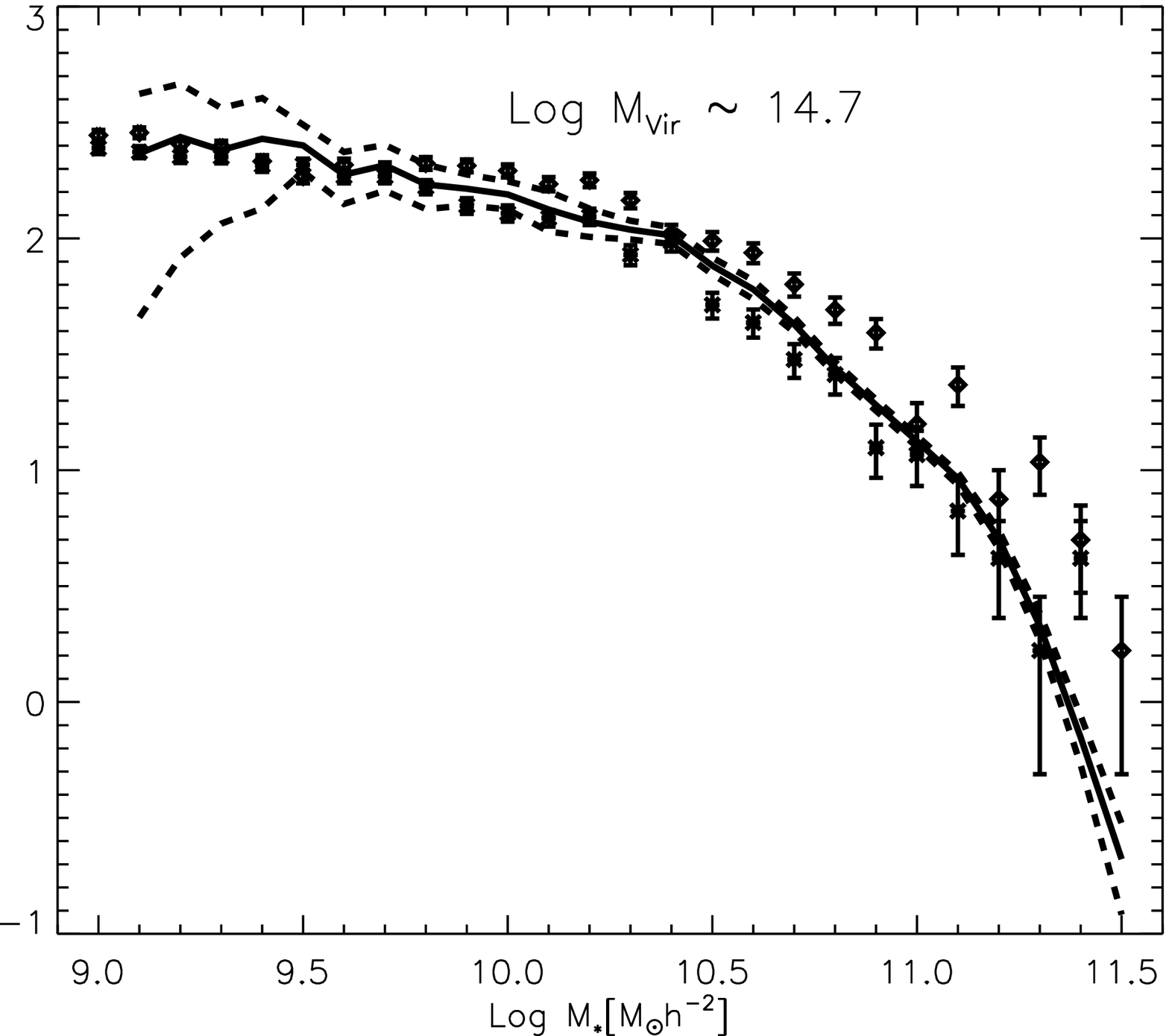} \\
\end{tabular}
\caption{Conditional SMF (CSMF) for satellites belonging to haloes of different mass (from $\log M_{Vir} \sim 12.9$ to $\log M_{Vir} \sim 14.7$) , as
predicted by Model D (stars), Model D with no stripping (diamonds), and compared with the observed abundance from SDSS data by \citet{yang09}. The solid
lines represent the observed data while dashed lines represent the scatter.}
\label{fig:CSMF}
\end{center}
\end{figure*}

In Figure \ref{fig:CSMF} we plot the Conditional SMF of satellite galaxies residing in haloes of different mass, from $\log M_{Vir} [M_{\odot}/h] \sim 12.9$ to 
$\log M_{Vir} [M_{\odot}/h] \sim 14.7$, as predicted by Model D (stars), and Model D with stripping switched off (diamonds), and compare our results with observations 
from SDSS survey (solid and dashed lines) by \citet{yang09}. A model with stellar stripping and a stripping efficiency (see Section \ref{sec:stripping}) halo mass 
dependent agrees better with observed data. From this figure it is also clear that a model with no stellar stripping would definitely over-predict the number density 
of satellites in groups over all the stellar mass range, and the number density of intermediate/massive satellites ($\log M_* [M_{\odot}h^{-2}] > 10$) in clusters.

It must be noted that the model as been tuned by using more halo bins than those showed in Figure A1, which shows only four examples. Altogether, halo bins used for 
the calibration of the model have a one-to-one correspondence with halo bins reported in Table \ref{tab:eta_values}. We assume that each value of $\eta$ found to match 
the Conditional SMF for each bin in Figure \ref{fig:CSMF} to be valid over a wider range in halo masses as those reported in Table \ref{tab:eta_values}.    

\section{Systematic uncertainties in stellar mass estimation}
\label{sec:append2}
Systematic uncertainties come as a consequence of different choices. Systematic differences in stellar masses due to SED fitting choices might be of the order of 0.2 dex 
(\citealt{santini14}). Such a difference in stellar mass is very important at the high-mass end because it results in a much larger difference in number density (with respect to the rest 
of the stellar mass range) due to the steepness of the SMF at high masses.  

Moreover, at low redshift the diffuse light around galaxies starts to be an important component (see, e.g., 
\citealt{giuseppe,myself2}) which might affect the high-mass end of the SMF depending whether it is detected or not. Our model predictions are not affected 
by this issue since the model is capable to separate the galaxy and its diffuse light (that is not considered in any plot of this paper). Nevertheless, the key-point concerning the 
detection of the ICL is model-independent and has non-negligible effects when comparing SMFs at different redshifts. In fact,  due to the cosmological surface brightness dimming, 
observations at high redshift likely do not include the contribution of the ICL in stellar mass measurements. According to \cite{behroozi13}, around 15\% of the stellar mass density 
at $z \sim 2$ is in ICL. This prediction seems to be extreme if compared to the results obtained by \cite{myself2} with a semi-analytic model of galaxy formation, but implies that a 
significant fraction of stellar mass of very massive galaxies at high redshift may not be measured. The caveat is necessary in the light of the fact that our model predictions at low
redshift are the result of the evolution of the observed SMF at high redshift that we use for tuning our model, which is affected by the systematic issue discussed above. In other words, 
a given amount of the mass evolution (likely negligible since observations typically are not capable to detect the diffuse light at high redshift) predicted by our model might not come 
from the actual growth of galaxies but from systematics in the measurements.

Other systematic differences can be due to a different choice of the IMF and to the stellar mass-to-light ratio (M/L). However, differences due to these two choices can be accounted 
for and corrected {\it a posteriori}.

\end{document}